\documentclass[]{gGAF2e}
\usepackage{multiequations}
\usepackage{amsmath}
\newcommand{\ovec}[1]{{\mbox{\boldmath $#1$}}}

\newcommand{\boldb}{\rm \textbf{b}}
\newcommand{\bB}{\rm \textbf{B}}

\newcommand{\be}{\rm \textbf{e}}

\newcommand{\bscE}{\ovec{\cal{E}}}

\newcommand{\kei}{\mbox{\textit{kei}}}
\newcommand{\bei}{\mbox{\textit{bei}}}
\newcommand{\ber}{\mbox{\textit{ber}}}

\newcommand{\bu}{\rm \textbf{u}}
\newcommand{\bU}{\rm \textbf{U}}

\newcommand{\mB}{\overline{B}}

\newcommand{\bQ}{\rm \textbf{Q}}

\newcommand{\bmB}{\overline{\rm \textbf{B}}}

\newcommand{\bmU}{\overline{\rm \textbf{U}}}
\newcommand{\balpha}{\ovec{\alpha}}

\newcommand{\bgamma}{\ovec{\gamma}}

\newcommand{\bnab}{\ovec{\nabla}}
\def\dd {\mbox{d}}

\def\x {\times}
\def\ol {\overline}
\def\lowap{\hbox{\space \raise-1mm\hbox{$ \stackrel{<}{\sim}$} \space}}
\def\greatap{\hbox{\space \raise-1mm\hbox{$ \stackrel{>}{\sim}$} \space}}

\begin{document}
\doi{10.1080/03091920xxxxxxxxx}
 \issn{1029-0419} \issnp{0309-1929} \jvol{00} \jnum{00} \jyear{2009} 

\markboth{Rossby waves and $\alpha$-effect}{Rossby waves and $\alpha$-effect}

\title{Rossby waves and $\alpha$-effect}

\author{R. AVALOS-ZUNIGA${\dag}$
\vspace{6pt}, F. PLUNIAN${\ddag}$$^{\ast}$\thanks{$^\ast$Corresponding author. Email: Franck.Plunian@ujf-grenoble.fr
\vspace{6pt}} and K. H. R\" ADLER${\dag\dag}$\\
\vspace{6pt}  ${\dag}$Universidad Aut\'onoma Metropolitana-Iztapalapa, Av. San Rafael Atlixco 186, col. Vicentina, 09340 D.F. M\'exico\\ 
${\ddag}$Laboratoire de G\'eophysique Interne et Tectonophysique, UJF, CNRS, BP 53, 38041 Grenoble Cedex 9, France\\
${\dag\dag}$Astrophysikalisches Institut Potsdam, An der Sternwarte 16,
D-14482 Potsdam, Germany\\
\vspace{6pt}\received{Received 6 September 2007; in final form 28 April 2009}}

\maketitle

\begin{abstract}
Rossby waves drifting in the azimuthal direction are a common feature at the onset of thermal convective instability in a rapidly rotating spherical shell. They can also result from the destabilization of a Stewartson shear layer produced by differential rotation as expected in the liquid sodium experiment (DTS) working in Grenoble, France.

A usual way to explain why Rossby waves can participate to the dynamo process goes back to \cite{Busse75}. In his picture, the flow geometry is a cylindrical array of parallel rolls aligned with the rotation axis. The axial flow component (the component parallel to the rotation axis) is (i) maximum in the middle of each roll and changes its sign from one roll to the next. It is produced by the Ekman pumping at the fluid containing shell boundary. The corresponding dynamo mechanism can be explained in terms of an $\alpha$-tensor with non-zero coefficients on the diagonal. It corresponds to the heuristic picture given by \cite{Busse75}.

In rapidly rotating objects like the Earth's core (or in a fast rotating experiment), Rossby waves occur in the limit of small Ekman number ($\approx 10^{-15}$). In that case, the main source of the axial flow component is not the Ekman pumping but rather the ``geometrical slope effect'' due to the spherical shape of the fluid containing shell. This implies that the axial flow component is (ii) maximum at the borders of the rolls and not at the centers. If assumed to be stationary, such rolls would lead to zero coefficients on the diagonal of the $\alpha$-tensor, making the dynamo probably less efficient if possible at all. Actually, the rolls are drifting as a wave, and we show that this drift implies  non--zero coefficients on the diagonal of the $\alpha$-tensor. These new coefficients are in essence very different from the ones obtained in case (i) and cannot be interpreted in terms of the heuristic picture of \cite{Busse75}. They were interpreted as higher-order effects in \cite{Busse75}. In addition we considered rolls not only drifting but also having an arbitrary radial phase shift as expected in real objects.\bigskip

\bigskip
\end{abstract}

\section{Introduction}
\label{introduction}
Rossby waves naturally result from thermal convection instabilities
in a rapidly rotating shell. Different configurations have been studied
depending on whether the fluid lies between two concentric spherical shells or inside a full sphere,
and on the type of heating (either differential or internal).
At the instability onset, the motion takes the form of rolls aligned with the axis of rotation and localized at the vicinity
of a cylinder lying in the bulk of the fluid. 
These rolls drift usually as a wave in the prograde azimuthal direction. 
In each roll in addition to the horizontal flow, that is, the flow in a plane
perpendicular to the axis of rotation,
there is also an axial flow (component parallel to the rotation axis)
due to the boundary conditions at the ends of the rolls.

In rapidly rotating objects like the Earth's core, the Ekman number $E$, defined as the ratio of the viscous to the Coriolis forces, is small ($E\propto 10^{-15}$).
Rossby wave as a linear solution of the thermal convection problem in the asymptotic limit of small Ekman number $E\ll 1$, was first proposed by \cite{Busse70} (see also the contribution by \citealt{Roberts68}).
After several intermediate improvements \citep{Soward77,Yano92}, 
an exact solution was given by \cite{Jones00}. Since then, the solution has been
confirmed numerically by \cite{Dormy04} (see also \citealt{Zhang91,Zhang92,Zhang93}).
These results assume the asymptotic limit $Pr/E\rightarrow \infty$ where $Pr$ is the Prandtl number defined as the ratio of the viscosity to the thermal diffusivity. Additional issues have been addressed,
in the asymptotic limit $Pr/E\rightarrow 0$ \citep{Zhang95} and in the general case $0\le Pr/E < \infty$ \citep{Zhang07}, including discussions about the nature of Rossby waves versus inertial waves \citep{Busse05}. 

Thermal Rossby waves have been studied also experimentally (\citealt{Busse76b,Carrigan83,Cardin94}, see also the review paper by \citealt{Busse02} and references therein). 
Above the onset the flow becomes highly turbulent and the non-linearities may be strong.
Though depending in a complex way on the parameters of the problem \citep{Grote01,Busse02,Morin04}, it is worth noting that
the persistence of the columnar structure of Rossby waves has been observed both
experimentally and numerically \citep{Aubert03,Cardin94,Sumita00}.

When the fluid is electrically conducting
such Rossby waves, in combination with differential rotation,
are expected to produce dynamo action \citep{Kageyama97}.
In addition, by processes related to a 2D inverse cascade \citep{Sommeria86,Aubert01} or to the presence of a strong toroidal magnetic field \citep{Cardin95}, 
the number of rolls for a very low Ekman number is expected to be much lower than estimated from the asymptotic theory of the onset of thermal convection.
Therefore, though the parameters in numerical
simulations or experiments are far from those of the Earth's outer core, these studies suggest that
the existence of Rossby waves in the form of columnar structures of reasonable size
may occur and be important in the geodynamo process.

Rossby waves can also be obtained mechanically, instead of thermally, 
as shown by \cite{Hide67} and \cite{Busse68}. More recently \cite{Schaeffer05a}
considered a fluid between two concentric spherical shells in fast rotation
but with slightly different rotation rates.  
They found that the destabilization of the Stewartson shear layer at the tangent cylinder leads indeed to Rossby waves. 
They have shown that such Rossby waves
with the strong differential rotation present in the fluid are capable of dynamo action \citep{Schaeffer06}. This has enforced the 
interest in building a new experiment in liquid sodium, called DTS \citep{Cardin02,Nataf06,Schmitt08}.

In dynamo theory, it is known for long that differential rotation, that is the
$\Omega$-effect, can generate a toroidal
magnetic field from a poloidal magnetic field (see e.g. \citealt{Elsasser56}).
The comprehensive picture given by \cite{Busse75}
illustrates how the interaction of Rossby waves with a toroidal magnetic field can generate
toroidal electric currents
and thus a poloidal magnetic field.
Within the mean--field concept this process can be described in terms of an $\alpha$--effect. 
The overall dynamo mechanism is known as an $\alpha\Omega$-dynamo.

\begin{figure}
\epsfig{file=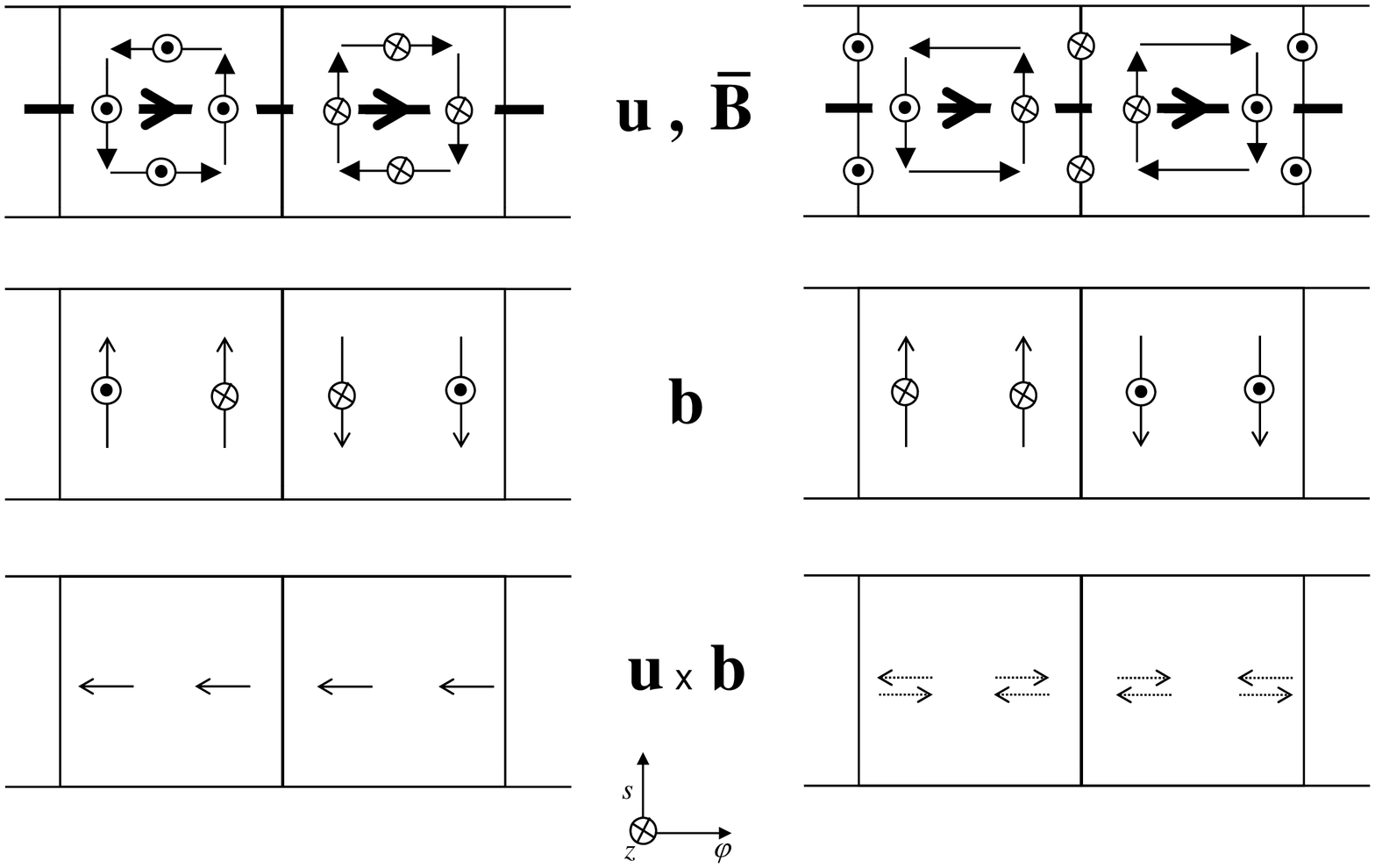,width=1\textwidth}\\*[-2cm]
\begin{center}
(i)\hspace{7cm}(ii)
\end{center}
\caption{In two adjacent rolls, the fluid velocity $\bu$ (small arrows and inwards/outwards symbols in first row) acting on a given azimuthal mean magnetic field $\bmB$ (thick arrow in top row) results
in a secondary magnetic field $\boldb$ (middle row), and so in an electromotive force $\bu \x \boldb$ (bottom row).
The radial, azimuthal and axial directions correspond to $s$, $\varphi$ and $z$.
Left column: case (i), for which the axial flow is maximum within each roll.
Right column: case (ii), for which the axial flow is maximum between each roll.
In case (i) $\bu$ and $\boldb$ are clearly not parallel or antiparallel,
and the azimuthal component of $\bu \x \boldb$ has the same sign everywhere in the considered points,
that is, its azimuthal average is non--zero.
By contrast in case (ii) $\bu$ and $\boldb$ are presumably more or less parallel or antiparallel
in the considered points so that $\bu \x \boldb$ is small.
Moreover the azimuthal component of $\bu \x \boldb$ has different signs in each cell.
These aspects suggest that its average vanishes.}
\label{heur}
\end{figure}

We can show by simple arguments,  that the efficiency of the process described by \cite{Busse75}, or the magnitude of the $\alpha$--effect,
depends critically on the relative positions of the horizontal and axial components of the flow
in each roll.
As a first step let us ignore the drift of the rolls.
As illustrated in figure \ref{fig:isopsic}a,
each roll rotates around its axis with a rotation rate changing its sign from
one roll to the adjacent ones.
We distinguish between the two cases in which the axial flow is maximum either (i) within each roll or (ii) between two adjacent rolls (top row of figure \ref{heur}).
A given large scale azimuthal magnetic field $\bmB$ is stretched by the fluid motion leading to a secondary magnetic field $\boldb$ (middle row of figure \ref{heur}). Then an azimuthal electromotive force $\bu \times \boldb$ is created (bottom row of figure \ref{heur}).
For case (i) all local electromotive forces
act in the same sense and so generate a global azimuthal electric current.
This can be interpreted as an $\alpha$--effect.
On the other hand, for case (ii) local electromotive forces with opposite signs
occur, implying eventually that there is no global azimuthal electric current, that is, no $\alpha$--effect.

In a rapidly rotating shell with a rigid boundary for which the no--slip condition applies,
and relying on the quasi-geostrophic approximation,
the axial flow $u_z$ in each roll is the sum of two terms.
One term is the Ekman pumping, scaling as $E^{1/2}$
and its intensity is maximum within each roll as in case (i).
The other term is the geometrical slope effect at the ends of the rolls, scaling as $E^{0}$ and
proportional to the radial flow $u_s$. It is 
then of maximum intensity between two adjacent cells as in case (ii).
It is argued in \cite{Schaeffer05a} that
in rapidly rotating spherical shells like the Earth's liquid core,
the Ekman number is so small that the first term in $u_z$ can be neglected
compared to the second one. Then in the light of the arguments
illustrated in figure \ref{heur}  (ii)
the ability of such a flow to generate an $\alpha$-effect for $E\ll 1$ could appear questionable.
Actually in such rapidly rotating systems, stationary convection can not occur. The rolls have to drift as a wave.
Then taking this drift into account, we shall demonstrate that even in case (ii) an $\alpha$--effect remains possible. A similar effect is described as an higher-order effect in \cite{Busse75}.
However we stress here that this $\alpha$--effect is in essence very different from the one that would be obtained 
with a flow geometry (i).

Finally, we stress that the $\alpha$-effect, so far understood as the generation of a toroidal mean electromotive force from a toroidal mean magnetic field, and described by only one $\alpha$-coefficient, is in fact a special case of a more general connection between mean electromotive force and mean magnetic field, described by an $\alpha$-tensor.
We shall calculate the additional coefficients of this $\alpha$-tensor and give their scaling properties in terms of the flow parameters like the number of rolls, the Rossby wave frequency and the magnetic Reynolds numbers (horizontal and vertical) of the flow. Depending on these parameters these additional coefficients may be dominant compared to the $\alpha$--effect mentioned above and then completely change the overall picture of the possible dynamo mechanism. 
We shall see that it is all the more true for rolls not only drifting but also having a radial shift as expected for Rossby waves traveling in a spherical shell. In that case the rolls are bent as illustrated in figure \ref{fig:isopsic}b.

In section \ref{sec:concept} we set the problem, give the basic equations, define the general
assumptions and specify the velocity field.
In section \ref{sec:calculation} we derive general analytical expressions for
the mean electromotive force with respect to a cylindrical coordinate system.
In section \ref{sec:example} we present asymptotic and numerical results for both cases (i) and (ii).
Finally, in section \ref{sec:Discussion} we discuss our results.

\section{General concept}
\label{sec:concept}
\subsection{Basic equations and assumptions}
We consider a body of a homogeneous electrically conducting incompressible fluid,
penetrated by a magnetic field $\bB$ and showing internal motions with a velocity $\bU$.
The magnetic field $\bB$ is assumed to be governed by the induction equation
\begin{multiequations}
\begin{equation}
\eta \bnab^2 \bB + \bnab \times (\bU \times \bB) - \partial_t \bB
= {\bf 0} \, , \quad \nabla \cdot \bB = 0,
\end{equation}
\label{eq:ind}
\end{multiequations}

where $\eta$ is the magnetic diffusivity, considered as constant.

Referring to a cylindrical coordinate system $(s, \varphi, z)$ we define mean fields
as in Braginskii's theory of the nearly axisymmetric dynamo \citep{Braginskii64a}
by averaging over $\varphi$.
Given a scalar field $F$, the corresponding mean field is denoted by $\ol{F}$.
It corresponds to the axisymmetric part of $F$.
In the case of vectors or tensors the same definition applies to each component so that,
e.g., $\bmB = (\ol{B}_s, \ol{B}_\varphi, \ol{B}_z)$.

We split $\bB$ and $\bU$ into mean fields, $\bmB$ and $\bmU$, and deviations $\boldb$ and $\bu$
from them, that is,
\begin{multiequations}
\begin{equation}
 \bB = \bmB + \boldb \, , \quad  \bU = \bmU + \bu \, .
\end{equation}
\label{eq:split}
\end{multiequations}
From the induction equation (\ref{eq:ind}) we obtain
\begin{multiequations}
\begin{equation}
\eta \bnab^2 \bmB +\bnab \times (\bmU \times \bmB)+\bnab \times \bscE - \partial_t \bmB
= {\bf 0} \, , \quad \nabla \cdot \bmB = 0,
\end{equation}
\label{eq:mind}
\end{multiequations}
where
\begin{equation}
\bscE =\overline{\bu \times \boldb}
\label{eq115}
\end{equation}
is the mean electromotive force due to $\bu$ and $\boldb$.

In view of the generation of a mean magnetic field, two terms of the equation (\ref{eq:mind})
are of particular interest, that with $\bmU$ and that with $\bscE$.
We assume here that the mean velocity $\bmU$ corresponds to a rotation only.
We further think of a proper specification of the small-scale velocity $\bu$
so that $\bscE$ covers the effect of Rossby waves. For the determination of
$\bscE$ we assume rigid-body mean rotation, that is we ignore any differential
rotation, and adopt a co-rotating frame of reference in which $\bmU=0$.
According to (\ref{eq115}) $\bscE$ is determined by $\bu$, which we consider as given, and $\boldb$.
Using (\ref{eq:ind}), (\ref{eq:split}) and (\ref{eq:mind}) we obtain
\begin{multiequations}
\begin{equation}
\eta \bnab^2 \boldb+\bnab \times (\bu \times \bmB)
+\bnab \times (\bu \times \boldb- \overline{\bu \times \boldb})  - \partial_t \boldb = {\bf 0} \, ,
\quad \nabla \cdot \boldb = 0 \, .
\end{equation}
\label{eq:find}
\end{multiequations}
Clearly this equation determines $\boldb$ if $\bu$ and $\bmB$ are given.

In order to make analytical calculations possible we introduce a quasi-linear 
approximation (also known as the second-order correlation approximation in mean field theory),
that is we neglect the term $|\bnab \times (\bu \times \boldb - \overline{\bu \times \boldb})|$
in (\ref{eq:find}). A sufficient condition for that approximation is
\begin{equation}
	\min (R_m', S_t) \ll 1
	\label{qla}
\end{equation}
with the magnetic Reynolds number $R_m'=u_0 l' / \eta$, and the Strouhal number $S_t = u_0 \tau / l'$, and
where $u_0$ is a typical magnitude of $\bu$,
$l'$ a characteristic small length scale of the roll and $\tau$ a characteristic time
of the Rossby wave. If $\tau$ is interpreted as the inverse wave frequency, $S_t$ is the ratio of
the roll's turn-over frequency to the wave frequency.
At the end of section \ref{sec:velocity} we will come back on the condition (\ref{qla}) for the applicability of the quasi-linear approximation.

We assume that the fluid velocity $\bu$ is non-zero only inside a cylindrical layer with the mean radius
$l_0$ and the thickness $2 \delta l_0$ ($\delta < 1$), that is, in $(1-\delta)l_0 \le s \le (1+\delta)l_0$. Its dependence on $s$ and $\varphi$ and on time will be specified later.
Moreover we consider $\bu$ as independent of $z$ (see e.g. \citealt{Kim99} for a similar approximation). Clearly $\bscE$
can only be different from zero inside that cylindrical layer.
As for $\bmB$, which is by definition independent of $\varphi$, we assume that it is independent on $z$, too. Further its time dependence is considered as weak compared to that of $\bu$ and therefore neglected in the following calculations. The independence of $\bu$ and $\bmB$ on $z$ suggests to consider also $\boldb$ as independent of $z$. Then also $\bscE$ does not longer depend on $z$.
Of course, in view of applications of our results to the Earth's core and the DTS experiment a treatment of the more general case with $z$-dependencies of $\bu$, $\bmB$, etc. would be of high interest. This is left for future work.

In what follows we measure all lengths in units of $l_0$,
the time in units of $l^2_0 / \eta$ and the velocity $\bu$ in units of $u_0$.
For several purposes it is useful to split $\bu$ into its parts
$\bu_\perp$ and $\bu_\parallel$ perpendicular and parallel to the rotation axis.
We measure $\bu_\perp$ and $\bu_\parallel$ in units of $u_{0\perp}$ and $u_{0\parallel}$
defined analogously to $u_0$.
With the assumptions introduced above, (\ref{eq:find}) turns then into
\begin{multiequations}
\begin{equation}
\bnab^2 \boldb - \partial_t \boldb = -\bnab \times \bQ \, , \quad
\bnab \cdot \boldb = 0
\end{equation}
\label{eq:B'}
\end{multiequations}
with
\begin{equation}
	\bQ = R_m \bu \times \bmB = (R_{m\,\perp} \bu_\perp + R_{m\,\parallel} \bu_\parallel) \x \bmB
\end{equation}
and with the magnetic Reynolds numbers
\begin{multiequations}
\begin{equation}
\trippleequation
R_m = \frac{u_0 l_0}{\eta} \, , \quad R_{m\,\perp} = \frac{u_{0\perp} l_0}{\eta} \, , \quad R_{m\,\parallel} = \frac{u_{0\parallel} l_0}{\eta}. 
\end{equation}
\label{eq:Rm}
\end{multiequations}
In contrast to $R_m'$, these quantities are defined with the large-scale parameter $l_0$.
\subsection{Fluid velocity}
\label{sec:velocity}
Since the fluid is considered as incompressible, we have $\bnab \cdot \bu = 0$.
As $\bu$ is taken independent of $z$,
we may represent it in the form
\begin{equation}
\bu = -\be \times \nabla \psi + u_z \be \, ,
\label{u1}
\end{equation}
where $\be$ is the unit vector in the $z$-direction. The stream function $\psi$ as well as $u_z$
may depend on $s$, $\varphi$ and $t$.

With the intention to simulate Rossby waves in their simplest form,
we further specify the velocity $\bu$ by
\begin{multiequations}
\begin{equation}
\psi = \tilde{\psi}(s) \, \cos(m(\varphi-\tilde{\varphi}(s)) - \omega t), \quad \quad \quad u_z=u_z(s) \cos(m(\varphi-\tilde{\gamma}(s)) - \omega t)
\end{equation}
\label{u2}
\end{multiequations}
with a positive integer $m$ and constant $\omega$.
The functions $\tilde{\varphi}(s)$ and $\tilde{\gamma}(s)$ are the radial dependent phase shifts of the horizontal and vertical flow components. They can be enforced for example by the spherical geometry. In this case they correspond to rolls with horizontal section shapes like bananas (see for example \citealt{Zhang07}). Two examples of the flow streamlines are given in figure \ref{fig:isopsic}. Each pattern drifts in the azimuthal direction with the dimensionless angular velocity $\omega / m$.

For later purposes we define a complex vector $\hat{\bu}(s)$,
depending on $s$ only, such that
\begin{equation}
\bu = {\rm Re}\{\hat{\bu}(s)\exp(-{\rm i}m\varphi + {\rm i}\omega t)\} \, .
\end{equation}
In a more explicit form, $\hat{\bu}(s)$ is given by
\begin{multiequations}
\singleequation
\begin{eqnarray}
\hat{u}_s &=& -\frac{{\rm i}m}{s} \tilde{\psi}(s) \exp ({\rm i}m\tilde{\varphi}(s)), \\ 
\hat{u}_{\varphi} &=& -[\partial_s \tilde{\psi}(s) +{\rm i}m\tilde{\psi}(s) \partial_s\tilde{\varphi}(s)] \exp ({\rm i}m\tilde{\varphi}(s)),\\
\hat{u}_z &=& u_z(s) \exp ({\rm i}m\tilde{\gamma}(s)). 
\end{eqnarray}
\label{velocity}
\end{multiequations}

Various relations between the horizontal and the axial flow, that is between $\psi$ and $u_z$,
can be specified by the choice of $\tilde{\varphi}(s)$ and $\tilde{\gamma}(s)$.
As we are mainly interested in the two cases (i) and (ii) described in figure \ref{heur},
we define the two corresponding flow types,
\begin{multiequations}
\begin{equation}
 \mbox{type (i)}: \tilde{\varphi}(s)=\tilde{\gamma}(s) \, , \quad 
\mbox{type (ii)} : \tilde{\varphi}(s)=\tilde{\gamma}(s)+\pi/2m.
\end{equation}
\label{flowtypes}
\end{multiequations}
The flow of type (i) corresponds to an axial flow driven by Ekman pumping.
The zero lines of the axial flow coincide then with the borders of the cells of the horizontal flows.
The flow of type (ii) corresponds to geometrical slope effect
as the main cause of the axial flow.
The extrema of the axial flow are then at the borders of the cells of the horizontal flow.

Returning to the condition (\ref{qla}) for the applicability of the quasi-linear approximation
we specify now $l'$ and $\tau$ such that $l'=\delta l_0$ and $\tau = l_0^2 / \eta \omega$. 
Then we have 
\begin{multiequations}
\begin{equation}
	R_m' = u_0 \delta l_0 / \eta \quad \quad \quad \quad \mbox{and} \quad \quad \quad S_t = u_0 l_0 / \eta \delta \omega = R_m' / \delta^2 \omega. 
\end{equation}
\end{multiequations}
For stationary flows ($\omega = 0$), irrelevant for Rossby waves but still of interest for comparison with some numerical simulations, the condition (\ref{qla}),
which is sufficient but not necessary for the validity of the quasilinear approximation, takes the form
$R_m' \ll 1$.
However, as turned out in simulations (Schrinner \textit{et al.} 2005, 2006), this approximation may well apply for values of $R_m'$ up to the order of unity.
For a non-stationary flow, if $\delta^2 \omega \gg 1$ the condition (\ref{qla}) turns into $S_t \ll 1$.
In view of the Earth's fluid core, let us consider that $l_0\approx 1800$ km and $\eta \approx 1$m$^2$s$^{-1}$
(which leads to $l_0^2/\eta \approx 10^5$ years). 
Assuming approximately equal extents of a roll in radial and azimuthal direction, $2 \delta = \pi/m$, and $m=16$ pairs of rolls we have $\delta \approx 0.1$ (and then a radius $\delta l_0$ of a roll about $180$ km). A typical drift velocity $\omega \eta / m l_0^2 = 0.2$ deg / year yields to $\omega \approx 5.6 \cdot 10^3$. Then we find that $\omega \delta^2 \approx 60$. Then again a sufficient condition for the applicability of the quasilinear approximation reads $S_t \ll 1$. It implies $u_0 \ll 0.3$ mm s$^{-1}$, which gives the flow intensity upper-limit above which the quasi-linear approximation might not work.

\section{Calculation of $\bscE$}
\label{sec:calculation}
\subsection{Poloidal-toroidal decomposition and reduction of equations}
In order to solve (\ref{eq:B'}), we represent $\boldb$
as a sum of poloidal and toroidal parts,
\begin{equation}
\boldb = - \bnab \times (\be \times \bnab S) - \be \times \bnab T
\label{eq125}
\end{equation}
with scalars  $S$ and $T$ depending  on $s$ and $\varphi$.
The components of $\boldb$ are then given by
\begin{multiequations}
\begin{equation}
\trippleequation
{b}_s = \frac{1}{s} \frac{\partial T}{\partial \varphi}, \, \quad
{b}_\varphi = - \frac{\partial T}{\partial s}, \, \quad
{b}_z = - D S  \, ,
\end{equation}
\label{eq127}
\end{multiequations}
where
\begin{equation}
D f = \frac{1}{s} \frac{\partial}{\partial s} \big( s \frac{\partial f}{\partial s} \big)
     + \frac{1}{s^2} \frac{\partial^2 f}{\partial \varphi^2} \, .
\label{eq1129}
\end{equation}
Likewise we represent $\bnab \times \bQ$ in the form
\begin{equation}
\bnab \times \bQ = \bnab \times (\be \times \bnab F) + \be \times \bnab G
\label{eq131}
\end{equation}
with scalars $F$ and $G$.
Using (\ref{eq:B'}), (\ref{eq125}) and (\ref{eq131})
we find, excluding singularities of $F$ and $G$ at $s = 0$ and for $s \to \infty$,
\begin{multiequations}
\begin{equation}
D S - \partial_t S= F \; , \;\; \;\;\; D T  - \partial_t T= G \, .
\end{equation}
\label{eq145}
\end{multiequations}
With the help of the identity $\be \cdot (\nabla \times (\be \times \nabla f)) = Df$
we further conclude from (\ref{eq131}) that
\begin{multiequations}
\begin{equation}
D F = \be \cdot (\bnab \times \bQ) \, , \quad
D G = \be \cdot (\bnab \times (\bnab \times \bQ))
    = - \be \cdot ( \bnab^2 \bQ - \bnab (\bnab \cdot \bQ)) \, .
\end{equation}
\label{eq133}
\end{multiequations}
The first of these relations can be written in the form
\begin{equation}
D F = \frac{1}{s} \big( \frac{\partial}{\partial s}(s Q_\varphi)
    - \frac{\partial Q_s}{\partial \varphi} \big) \, .
\label{eq135}
\end{equation}
The second one is equivalent to $D G = - D Q_z$ or, if we exclude again singularities
at $s = 0$ and for $s \to \infty$, to
\begin{equation}
G=-Q_z \, .
\label{eq:GQz}
\end{equation}

\subsection{Calculation of $\boldb$}
Like the components of $\bu$, those of $\bQ$
as well as the functions $F$ and $G$ have the form
\begin{equation}
f(s, \varphi) = f^{(c)} (s) \cos (m \varphi - \omega t) + f^{(s)} (s) \sin (m \varphi - \omega t)
= {\rm Re}\{\hat{f}(s)\exp(-{\rm i}m\varphi +{\rm i} \omega t)\} \, ,
\label{eq163}
\end{equation}
where $\hat{f}$ is a complex quantity depending on $s$ only.
The same applies to $\boldb$ as well as $S$ and $T$.
In this notation the relations (\ref{eq127}) take the form
\begin{multiequations}
\begin{equation}
\trippleequation
\hat{b}_s = - \frac{{\rm i}m}{s} \, \hat{T} (s) \, , \quad
\hat{b}_\varphi = - \frac{\dd \hat{T}(s)}{\dd s} \, , \quad
\hat{b}_z =  -D_m \hat{S}(s)
\end{equation}
\label{eq205}
\end{multiequations}
with
\begin{equation}
D_m f = \frac{1}{s} \frac{\partial}{\partial s} \big( s \frac{\partial f}{\partial s} \big)
    - \frac{m^2}{s^2} f \, .
\label{eq159}
\end{equation}
The equations (\ref{eq145}), (\ref{eq135}) and (\ref{eq:GQz}) reduce to
\begin{multiequations}
\begin{equation}
(D_m  - {\rm i} \omega) \hat{S} = \hat{F} \, , \quad 
(D_m  - {\rm i} \omega) \hat{T} = \hat{G} \, ,
\end{equation}
\label{eq203}
\end{multiequations}
\begin{multiequations}
\begin{equation}
D_m \hat{F} = \frac{1}{s} \big( \frac{\partial}{\partial s}(s \hat{Q}_\varphi)
   +{\rm i}m \hat{Q}_s \big), \quad \hat{G} = -\hat{Q}_z.
\end{equation}
\label{eq165}
\end{multiequations}
In order to determine $\hat{\boldb}$
we have to solve (\ref{eq203}) with $\hat{F}$ and $\hat{G}$ satisfying (\ref{eq165}).
Generic solutions of such equations are derived in Appendix \ref{mathematics}.
According to (\ref{eq1713})-(\ref{eq1743}) the solution of the first equation of (\ref{eq165}) can be written in the form
\begin{equation}
\hat{F}(s)= - \int_0^\infty \frac{h_m (s, s')}{s'}
   \big(\frac{\partial}{\partial s'}(s' \hat{Q}_\varphi (s'))
   + {\rm i}m \hat{Q}_s (s') \big) \, s' \dd s',
\label{eq2083}
\end{equation}
where $h_m (s, s')$ is the Green's function defined in (\ref{eq1753}).
After integrating by parts, we have
\begin{equation}
\hat{F} (s) =  \int_0^\infty
   \big(\frac{\partial h_m (s, s')}{\partial s'} \, \hat{Q}_\varphi (s')
   -{\rm i}m \frac{h_m (s, s')}{s'} \, \hat{Q}_s (s') \big) \, s' \, \dd s' \, .
\label{eq2093}
\end{equation}
Now from (\ref{eq165}), (\ref{eq2093}) and with the help of (\ref{eq171})-(\ref{eq260}), the solutions of (\ref{eq203}) turn into
\begin{multiequations}
\singleequation
\begin{eqnarray}
\hat{S} (s) &=& -\int_{0}^\infty \int_0^\infty
    k_m(s,s')\big(\frac{\partial h_m}{\partial s''}(s',s'') \,
    \hat{Q}_\varphi (s'')
    - \frac{{\rm i}m}{s''}h_m(s',s'') \,
    \hat{Q}_s (s'') \big) \,
    s'' \dd s'' s' \dd s',\\
\hat{T}(s) &=& \int_0^\infty k_m (s,s') \, \hat{Q}_z (s') \, s' \, \dd s',
\end{eqnarray}
\label{ST}
\end{multiequations}
where $k_m(s,s')$ is the Green's function defined in (\ref{eq261append2}).\\

Let us return to $\hat{\boldb}$ as given by (\ref{eq205}).
With (\ref{ST}) and (\ref{eq203}) we obtain
\begin{multiequations}
\singleequation
\begin{eqnarray}
\hat{b}_s (s) &=& - \int_0^\infty \frac{{\rm i}m}{s}k_m (s,s') \, \hat{Q}_z (s') \, s' \, \dd s',\\
\hat{b}_\varphi (s) &=& - \int_0^\infty \frac{\partial k_m}{\partial s}(s,s') \, \hat{Q}_z (s') \, s' \, \dd s',\\
\hat{b}_z (s) &=& - \int_0^\infty
   \big(\frac{\partial h_m (s, s')}{\partial s'} \, \hat{Q}_\varphi (s')
   -{\rm i}m \frac{h_m (s, s')}{s'} \, \hat{Q}_s (s') \big) \, s' \, \dd s' \nonumber\\
   &+& {\rm i} \omega \int_{0}^\infty \int_0^\infty
    k_m(s,s')\big(\frac{\partial h_m}{\partial s''}(s',s'') \,
    \hat{Q}_\varphi (s'')
    - \frac{{\rm i}m}{s''}h_m(s',s'') \,
    \hat{Q}_s (s'') \big) \,
    s'' \dd s'' s' \dd s'.
\end{eqnarray}
\label{eq293}
\end{multiequations}
After some algebra we further find
\begin{equation}
\hat{b}_z (s) = \int_0^\infty
   \big [ \big(-\frac{\partial h_m (s, s')}{\partial s'}+ {\rm i} \omega A_1(s,s') \big) \hat{Q}_\varphi (s') 
   + \frac{{\rm i}m}{s'} \big(h_m (s, s')-{\rm i} \omega A_2(s,s')\big) \hat{Q}_s (s') \big ] \, s' \, \dd s'
   \label{eqbz}
\end{equation}
with
\begin{multiequations}
\begin{equation}
	A_1(s,s')= \int_0^\infty k_m(s,s'') \frac{\partial h_m (s'', s')}{\partial s'} s'' \dd s'', \quad
	A_2(s,s')= \int_0^\infty k_m(s, s'') h_m (s'', s') s'' \dd s''.
\end{equation}
\end{multiequations}
With the help of relations of Appendix \ref{sec:Bessel} we can show that
\begin{multiequations}
\begin{equation}
	A_1(s,s')= \frac{1}{{\rm i}\omega}\frac{\partial}{\partial s'} (h_m - k_m)(s,s') \; , \quad
	A_2(s,s')= \frac{1}{{\rm i}\omega}(h_m - k_m)(s,s').
\end{equation}
\end{multiequations}
This leads to
\begin{eqnarray}
\hat{b}_z (s) = \int_0^\infty
   \big [ -\frac{\partial k_m (s, s')}{\partial s'} \hat{Q}_\varphi (s') 
   + \frac{{\rm i}m}{s'} k_m (s, s') \hat{Q}_s (s') \big ] \, s' \, \dd s'.
   \label{eqbz2}
\end{eqnarray}

From the second relation (\ref{eq:B'}) we have
\begin{multiequations}
\singleequation
\begin{eqnarray}
\hat{Q}_s &=&     R_{m\,\perp}\hat{u}_\varphi \mB_z - R_{m\,\parallel}\hat{u}_z \mB_\varphi,\\
\hat{Q}_\varphi &=& R_{m\,\parallel}\hat{u}_z \mB_s - R_{m\,\perp}\hat{u}_s \mB_z,\\
\hat{Q}_z &=&       R_{m\,\perp} \big(\hat{u}_s \mB_\varphi - \hat{u}_\varphi \mB_s\big) \, .
\end{eqnarray}
\label{eq167}
\end{multiequations}
Inserting this into (\ref{eq293}) and (\ref{eqbz2}) we obtain
\begin{multiequations}
\singleequation
\begin{eqnarray}
\hat{b}_s (s) &=& -{\rm i}m \, R_{m\,\perp} \, \int_0^\infty \frac{k_m (s, s')}{s} \,
     \big( \hat{u}_s (s') \, \mB_\varphi (s') - \hat{u}_\varphi (s') \, \mB_s (s') \big) \, s' \, \dd s',\\
\hat{b}_\varphi (s) &=& -R_{m\,\perp} \,
     \int_0^\infty \frac{\partial k_m (s, s')}{\partial s} \,
     \big(\hat{u}_s (s') \, \mB_\varphi (s') - \hat{u}_\varphi (s') \, \mB_s (s')\big) \, s' \, \dd s',\\
\hat{b}_z (s) &=& - R_{m\,\parallel} \,
     \int_0^\infty  \, \big[ \frac{\partial k_m (s, s')}{\partial s'} \, \mB_s (s')
     +\frac{{\rm i}m}{s'} k_m (s, s')\mB_\varphi (s') \big]\hat{u}_z(s') \, s' \, \dd s'\nonumber \\
&+& R_{m\,\perp} \,
     \int_0^\infty \big[ \frac{\partial k_m (s, s')}{\partial s'} \,
     \hat{u}_s (s')
     +\frac{{\rm i}m}{s'} k_m (s, s') \hat{u}_\varphi (s') \big] \mB_z (s') \, s' \, \dd s' \,.
\end{eqnarray}
\label{eq211}
\end{multiequations}
As mentioned in Appendix \ref{mathematics}, in the stationary case, that is $\omega = 0$, $k_m$ turns into $h_m$.
We note that as the fluid is at rest outside the moving layer, the integrations in (\ref{eq211}) and previous equations can be reduced to $1-\delta \le s \le 1+\delta$.

\subsection{Integral representation of $\bscE$}
\label{subsres}
According to (\ref{eq115}) we have
\begin{equation}
	{\bscE} = \frac{u_0}{2} {\rm Re} \left\{ \hat{\bu}^{*}\times \hat{\boldb} \right\},
	\label{eq181a}
\end{equation} 
where $\hat{\bu}^{*}$ is the complex conjugate of $\hat{\bu}$.
When inserting $\hat{b}_s (s), \hat{b}_{\varphi} (s), \hat{b}_z (s)$
as given by (\ref{eq211}) we see that $\bscE$ can be written in the form
\begin{equation}
{\cal{E}}_\kappa (s) = \int_{1-\delta}^{1+\delta}
    K_{\kappa \lambda} (s, s') \, \mB_\lambda (s') \, s' \, \dd s'.
\label{eq183}
\end{equation}
Here and in what follows $\kappa$ and $\lambda$ stand for $s$, $\varphi$ or $z$.
The kernel $K_{\kappa \lambda}$ is then given by
\begin{equation}
K_{\kappa \lambda} = \frac{\eta}{l_0} R_{m \perp}
\left\{ \begin{array}{c}
\!\!\! R_{m \perp} \!\!\! \\
\!\!\! R_{m \parallel} \!\!\! \\
\end{array} \right\} \,
{\tilde{K}}_{\kappa \lambda} \;\; \mbox{if} \;\;
(\kappa \lambda) = \left\{ \begin{array}{cccc}
\!\!(s \, z),&\!\!(\varphi \, z),&\!\!(z \, s),&\!\!(z \, \varphi)\\
\!\!(s \, s),&\!\!(s \, \varphi),&\!\!(\varphi \, s),&\!\!(\varphi \, \varphi)\\
\end{array} 
\right.
\, , \;\; K_{z z} = 0 \, .
\label{eq:kappaHZ}
\end{equation}
The dimensionless quantities ${\tilde{K}}_{\kappa \lambda}$, which do no longer depend on $R_{m \perp}$
or $R_{m \parallel}$, are given by
\begin{multiequations}
\singleequation
\begin{eqnarray}
4\tilde{K}_{s s} (s, s')&=& -
      \frac{\partial k_m }{\partial s'}(s, s') \, \hat{u}^{*}_\varphi (s)
     \, \hat{u}_z (s')
     - \frac{\partial k_m }{\partial s}(s, s') \hat{u}_z^{*} (s)
     \, \hat{u}_\varphi (s') +{\rm c.c.},\\
4\tilde{K}_{s \varphi} (s, s')&=& -
      \frac{{\rm i}m}{s'} k_m (s, s') \, \hat{u}^{*}_\varphi (s)
     \, \hat{u}_z (s')
     + \frac{\partial k_m }{\partial s}(s, s') \hat{u}_z^{*} (s)
     \, \hat{u}_s (s') +{\rm c.c.},\\
4\tilde{K}_{s z} (s, s')&=&
      +\frac{\partial k_m }{\partial s'}(s, s') \, \hat{u}^{*}_\varphi (s)
     \, \hat{u}_s (s')
     + \frac{{\rm i}m}{s'} k_m (s, s')\,\hat{u}^{*}_\varphi (s)\, \hat{u}_\varphi (s') +{\rm c.c.},\\
4\tilde{K}_{\varphi s} (s, s')&=&
     + \frac{{\rm i}m}{s} k_m (s, s') \, \hat{u}_z^{*} (s)
     \, \hat{u}_\varphi (s')
     + \frac{\partial k_m }{\partial s'}(s, s')\, \hat{u}^{*}_s (s)
     \, \hat{u}_z (s') +{\rm c.c.}, \\
4\tilde{K}_{\varphi \varphi} (s, s')  &=&  
     -\frac{{\rm i}m}{s} k_m (s, s') \, \hat{u}_z^{*} (s) \, \hat{u}_s (s')
     + \frac{{\rm i}m}{s'} k_m (s, s') \, \hat{u}^{*}_s (s) \, \hat{u}_z (s') +{\rm c.c.}, \\
4\tilde{K}_{\varphi z} (s, s')  &=&  -
      \frac{\partial k_m }{\partial s'}(s, s') \, \hat{u}^{*}_s (s) \, \hat{u}_s (s')
    - \frac{{\rm i}m}{s'} k_m (s, s')\hat{u}^{*}_s (s) \, \hat{u}_\varphi (s') +{\rm c.c.},\\
4\tilde{K}_{z s} (s, s')  &=&
     +\frac{\partial k_m }{\partial s}(s, s')
\hat{u}^{*}_s (s) \, \hat{u}_\varphi (s') 
    -\frac{{\rm i}m}{s} k_m (s, s')\,\hat{u}^{*}_\varphi (s) \, \hat{u}_\varphi (s') +{\rm c.c.},\\
4\tilde{K}_{z \varphi} (s, s')  &=&  -
     \frac{\partial k_m }{\partial s}(s, s')
 \hat{u}^{*}_s (s) \, \hat{u}_s (s') + \frac{{\rm i}m}{s}k_m (s, s') \,\hat{u}^{*}_\varphi (s) \, \hat{u}_s (s')+{\rm c.c.},
\end{eqnarray}
\label{eq214b}
\end{multiequations}
where ${\rm c.c.}$ denotes the complex conjugation.
By symmetry reasons and because of the $z$-independence of the flow we always have $K_{zz}=0$.

\subsection{Expansion of $\bscE$}
\label{sub:coeff}
We rely now on the integral representation (\ref{eq183}) of ${\cal{E}}_\kappa$,
assume that $\mB_\lambda (s)$ varies only weakly with $s$ and use the Taylor expansion
\begin{equation}
\mB_\lambda (s') = \mB_\lambda (s)
+ (s' - s) \, \frac{\partial \mB_\lambda (s)}{\partial s} + \cdots \, .
\label{eq185}
\end{equation}
In this way we obtain
\begin{equation}
{\cal{E}}_\kappa (s) = a_{\kappa \lambda} (s) \, \mB_\lambda (s)
   + b_{\kappa \lambda s} (s) \,
   \frac{\partial \mB_\lambda (s)}{l_0 \partial s}
   + \cdots
\label{eq187}
\end{equation}
with
\begin{multiequations}
\begin{equation}
a_{\kappa \lambda} (s) = \int_{1-\delta}^{1+\delta}
   K_{\kappa \lambda} (s, s') \, s' \, \dd s',
\quad \quad \quad
b_{\kappa \lambda s} (s) = l_0 \,\int_{1-\delta}^{1+\delta}
   K_{\kappa \lambda} (s, s') \, (s'-s) \, s' \, \dd s'.
\end{equation}
\label{eq189}
\end{multiequations}
The factors $l_0$ have been inserted in (\ref{eq187}) and (\ref{eq189}) in order to give $b_{\kappa \lambda s}$
the  dimension of a magnetic diffusivity.
The terms with higher derivatives of $\mB_\lambda$,
indicated by $\cdots\,$, are ignored in the following.

Expressing ${K}_{\kappa \lambda}$ in the relations (\ref{eq189})
in terms of ${\tilde{K}}_{\kappa \lambda}$
we obtain
\begin{equation}
a_{\kappa \lambda} = \frac{\eta}{l_0} R_{m \perp}
\left\{ \begin{array}{c}
\!\!\! R_{m \perp} \!\!\! \\
\!\!\! R_{m \parallel} \!\!\! \\
\end{array} \right\} \,
{\tilde{a}}_{\kappa \lambda} \;\; \mbox{if} \;\;
(\kappa \lambda) = \left\{ \begin{array}{cccc}
\!\!(z \, s),&\!\!(z \, \varphi)&&\\
\!\!(s \, s),&\!\!(s \, \varphi),&\!\!(\varphi \, s),&\!\!(\varphi \, \varphi)\\
\end{array} \right. \, , \quad
a_{\kappa z} = 0,
\label{eq:abHZ}
\end{equation}
\begin{equation}
b_{\kappa \lambda s} = \eta R_{m \perp}
\left\{ \begin{array}{c}
\!\!\! R_{m \perp} \!\!\! \\
\!\!\! R_{m \parallel} \!\!\! \\
\end{array} \right\} \,
{\tilde{b}}_{\kappa \lambda s} \;\; \mbox{if} \;\;
(\kappa \lambda) = \left\{ \begin{array}{cccc}
\!\!(s \, z),&\!\!(\varphi \, z),&\!\!(z \, s),&\!\!(z \, \varphi)\\
\!\!(s \, s),&\!\!(s \, \varphi),&\!\!(\varphi \, s),&\!\!(\varphi \, \varphi)\\
\end{array} \right. \, , \quad
b_{z z s} = 0,
\label{eq:bbHZ}
\end{equation}
where 
${\tilde{a}}_{\kappa \lambda}$
and
${\tilde{b}}_{\kappa \lambda s}$ are dimensionless quantities independent of 
$R_{m \perp}$ and $R_{m \parallel}$.
Analytical expressions for the $\tilde{a}_{\kappa \lambda}$
and $\tilde{b}_{\kappa \lambda s}$ can be derived 
on the basis of (\ref{eq:kappaHZ}), (\ref{eq214b}) and (\ref{eq189}).
Of course, $a_{zz}=0$ and $b_{zzs}=0$ follow from $K_{zz}=0$.
As for $a_{sz}=a_{\varphi z} = 0$ we note that 
$\int_{1-\delta}^{1+\delta} K_{sz}(s,s')s' ds'$ and $\int_{1-\delta}^{1+\delta} K_{\varphi z}(s,s')s' ds'$ with 
$K_{s z}$ and $K_{\varphi z}$ according to (\ref{eq214b}) vanish. This can be shown with integrations by part and using $\bnab \cdot \bu = 0$.

Let us add a remark on the nature of the expansion (\ref{eq187}) and the coefficients $a_{\kappa \lambda}$
and $b_{\kappa \lambda s}$.
In most representations of mean--field electrodynamics the connection between
$\bscE$, $\bmB$ and its derivatives is, with respect to a Cartesian coordinate system, given in the form
${\cal{E}}_i = \breve{a}_{ij} \mB_j + \breve{b}_{ijk} \partial \mB_j / \partial x_k + \cdots$.
(Usually the notation $a_{ij}$ and $b_{ijk}$ is used instead of $\breve{a}_{ij}$ and $\breve{b}_{ijk}$.
We deviate from that, since $a_{ij}$ and $b_{ijk}$ are already otherwise defined in this paper.)
It is understood as a coordinate--independent connection, which implies that
${\cal{E}}_i$ and $\mB_j$ are components of vectors,
and $\breve{a}_{ij}$, $\breve{b}_{ijk}$ as well as $\partial \mB_j / \partial x_k$ components of tensors,
all with the well--known behavior of such objects under coordinate transformations.
The $a_{\kappa \lambda}$, however, do not completely coincide
with the components of the tensor derived in that sense from the $\breve{a}_{ij}$.
The reason is that the transformation of $\partial \mB_j / \partial x_k$
in our cylindrical coordinate system
produces not only terms with derivatives of $\mB_\kappa$ but also such without derivatives
(the same remark would apply if our coordinate system was spherical instead of being cylindrical).
In the common understanding the $\alpha$--effect is, again in coordinate--independent manner,
defined on the basis of the contribution $\breve{a}_{ij} \mB_j$ to ${\cal{E}}_i$.
We slightly deviate from this definition in what follows.
When speaking of $\alpha$--effect we refer simply
to the contribution $a_{\kappa \lambda} \mB_\lambda$ to ${\cal{E}}_\kappa$.
This is in so far justified as
it is just this contribution which describes, e.g., the generation of ${\cal{E}}_\varphi$
from $\mB_\varphi$.

\section{Some typical examples }
\label{sec:example}
\subsection{Specification of the flow patterns}
\label{sec:flow patterns}
\begin{figure}
\begin{center}
\epsfig{file=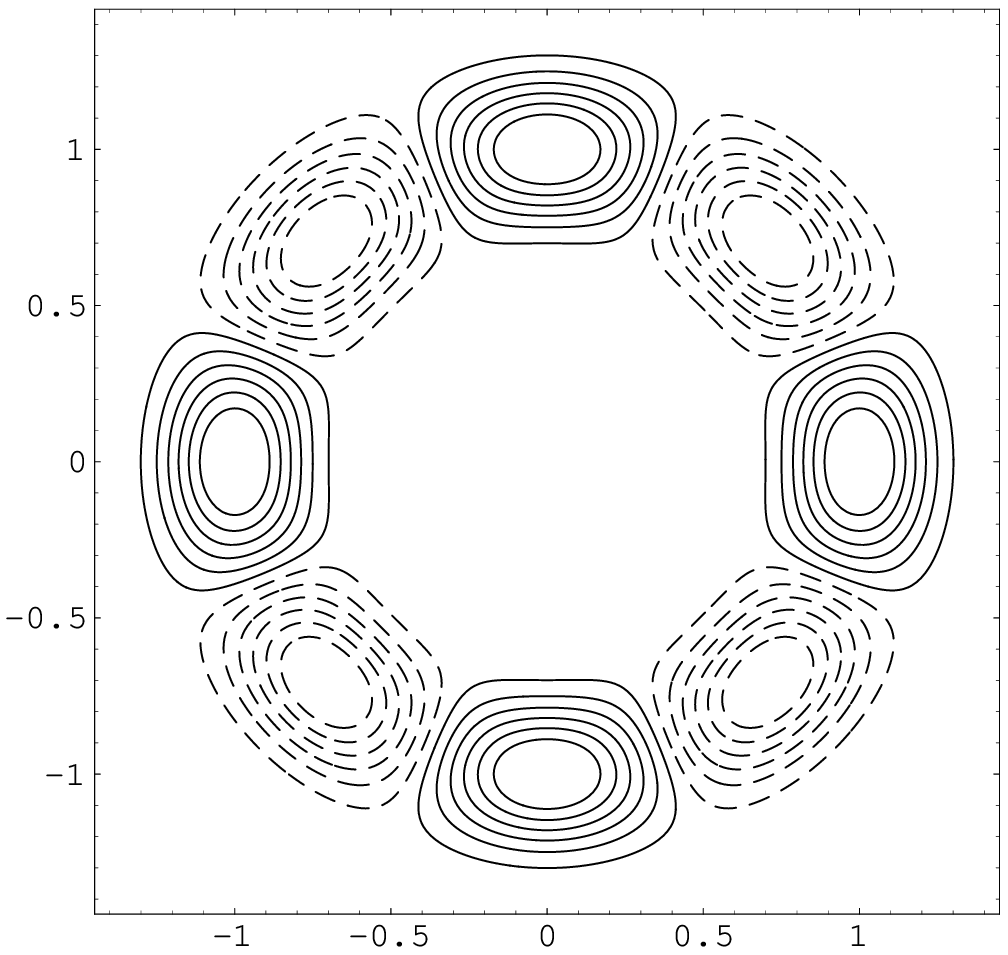,width=0.32\textwidth}
\epsfig{file=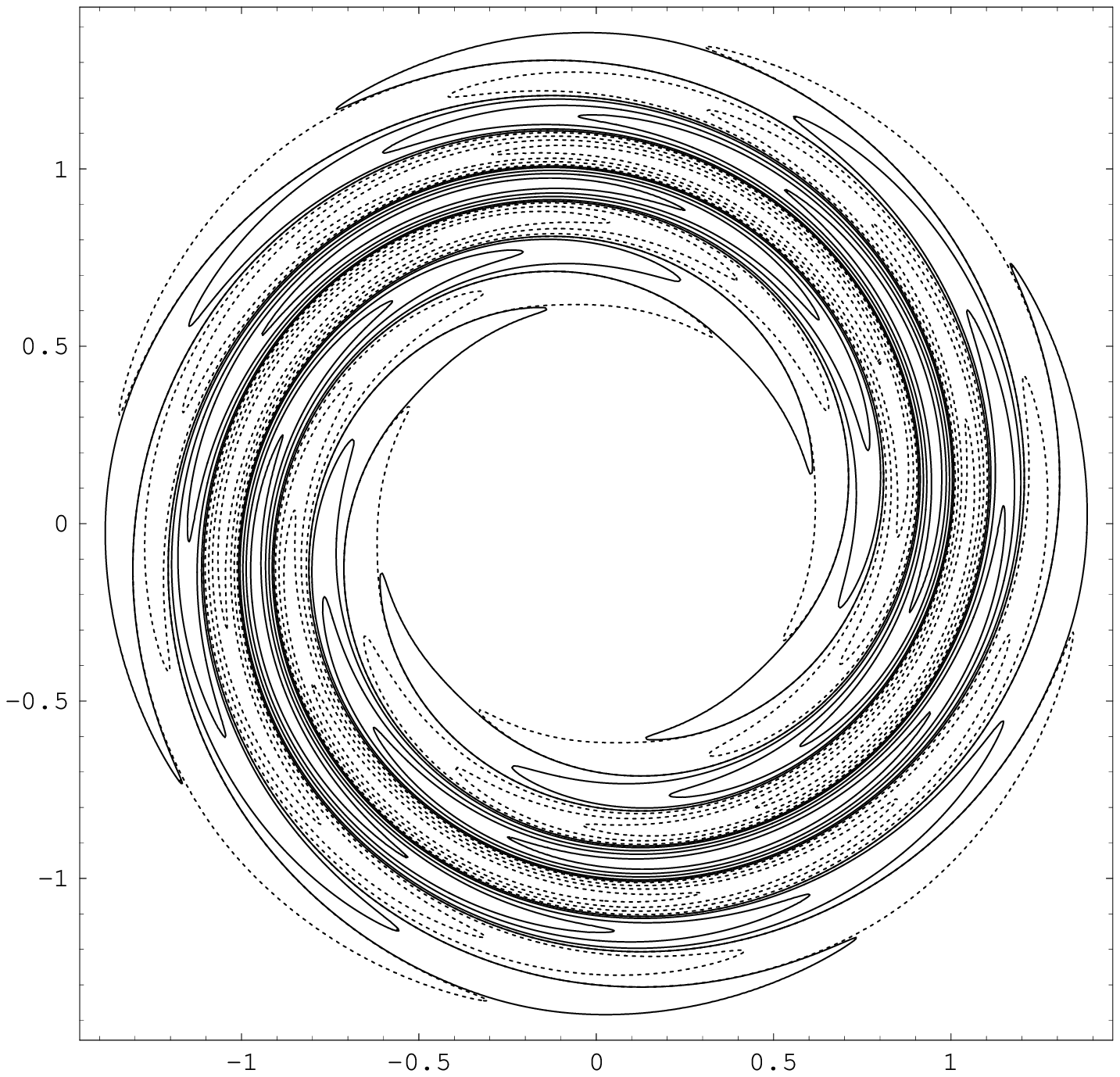,width=0.32\textwidth}
\caption{Isolines of $u_z$ as defined by (\ref{u2}) and according to (\ref{eq227b}) for $\delta m = \pi / 2$ and $m=4$.
Solid and dashed lines indicate opposite circulations. The left and right figures correspond to cases (a) and (b) given in (\ref{shifts}).}
\label{fig:isopsic}
\end{center}
\end{figure}

In this section we present numerical results
for some typical flow profiles corresponding to rolls of type (i) and  
(ii) as defined in (\ref{flowtypes}). 
In addition we also consider two types of phase-shift radial dependence,
\begin{multiequations}
\begin{equation}
	(a) \quad \tilde{\gamma}(s) = 0 \quad \quad \mbox{and} \quad \quad (b) \quad  \tilde{\gamma}(s) = \frac{\pi(s+\delta-1)}{\delta}.
\end{equation}
\label{shifts}
\end{multiequations}
The flow geometry is further specified by
\begin{eqnarray}
&& u_z = \frac{15 \pi}{16} (1 - \xi^2)^2 \, , \quad
\tilde{\psi} = \delta (1 - \xi^2)^3 \, , \quad
\xi = \frac{s-1}{\delta} \, , \quad \mbox{if} \quad |\xi| < 1,
\nonumber\\
&& u_z = \tilde{\psi} =0 \quad \mbox{otherwise} \, .
\label{eq227b}
\end{eqnarray}
Here $u_z$ and $\tilde{\psi}$ are normalized such that at any time the average of $u_z$
over a surface given by $1 - \delta \leq s \leq 1 + \delta$ and
$- \pi/2m \leq \varphi \leq \pi/2m$
as well as the average of $u_\varphi$ at a given value of $\varphi$
over $1 \leq s \leq 1 + \delta$ are equal to unity.
Figure \ref{fig:isopsic} shows isolines of $u_z$ for the two case (a) and (b) defined in (\ref{shifts}).

In case (a) the extent of a flow cell is $2 \delta$ (in units of $l_0$) in the radial direction and $\pi / m$
in the azimuthal direction.
We speak of ``compact rolls" if their ratio $2 \delta m / \pi$, or simply $\delta m$,
is in the order of unity. In that sense Fig.~\ref{fig:isopsic}a shows compact rolls.
In case (b) the radial phase-shift leads to extended rolls as shown in Fig.~\ref{fig:isopsic}b
even if $\delta m$ is in the order of unity. For simplicity we always set $2 \delta m / \pi=1$
in the rest of the paper.
\subsection{A first analysis of the results}
\label{first analysis}
The results obtained so far allow us to draw some conclusions
concerning the structure of $\bscE$ and the $\alpha$--effect.
Let us consider the kernel $K$ for flows of types (i) and (ii)
defined in (\ref{flowtypes})
and recall that their axial parts are driven by Ekman pumping or the geometrical slope effect, respectively.\\
\\
In the stationary case ($\omega = 0$), irrelevant for Rossby waves but still of general interest,
with the help of (\ref{flowtypes}) and (\ref{eq214b}), we conclude that $K$ has the following structure
\begin{multiequations}
\begin{equation}
K = \left( \begin{array}{ccc}
\times&0&0\\
0&\times&\times\\
0&\times&0
\end{array} \right) \; \, \mbox{in case (i)} \, , \quad
K = \left( \begin{array}{ccc}
0&\times&0\\
\times&0&\times\\
0&\times&0
\end{array} \right) \; \, \mbox{in case (ii)} \, ,
\end{equation}
\label{kernel}
\end{multiequations}
where crosses stand for matrix elements which are not necessarily equal to zero.
In case (i) we see from (\ref{kernel}) that $K$ leads to non-zero coefficients $a_{ss}$ and $a_{\varphi \varphi}$
corresponding to the dynamo mechanism described by \cite{Busse75} which can be interpreted as an
$\alpha^2$-mechanism. 
In the Karlsruhe experiment (\citealt{Muller04,Muller06}) the dynamo action of a mechanism
of that kind has been demonstrated.
On the other hand, in case (ii) (\ref{kernel}) implies that $a_{\varphi \varphi}=0$ (and $a_{ss}=0$) in accordance with the heuristic arguments of figure \ref{heur}.\\
\\   
For drifting waves ($\omega \ne 0$), the kernel components except $K_{zz}$ are not
necessary equal to zero. 
Then it may happen that the coefficients off-diagonal be dominant and therefore the simple $\alpha$-effect mentioned earlier not be a relevant part of the dynamo mechanism.
In fact, in addition to the $\alpha$-effect a transport of mean magnetic flux by the so-called $\gamma$-effect can be expected. To derive it
we can define a symmetric matrix $\balpha$ and a vector $\bgamma$
in the following way
\begin{multiequations}
\begin{equation}
	\alpha_{\kappa \lambda} = - \frac{1}{2}(a_{\kappa \lambda} + a_{\lambda \kappa})
	\quad,  \quad \quad \quad \gamma{\kappa} = \frac{1}{2} \epsilon_{\kappa \lambda \mu} a_{\lambda \mu},
\end{equation}
\end{multiequations}
where $\epsilon_{\kappa \lambda \mu}$ has to be interpreted in the usual way (Levi-Civita symbol)
identifying the subscripts $s$, $\varphi$ and $z$ with 1, 2 and 3, respectively.
Then we can write the mean electromotive force in the form
\begin{equation}
\bscE = - \balpha \circ \bmB - \bgamma \times \bmB
\end{equation}
in which the $s$-derivatives of $\bmB$ are not included.
As can be seen from the structures of the matrix $[a_{\kappa \lambda}]$ 
contributions to both $\balpha$ and $\bgamma$ are expected.
If the role of the coefficient $\alpha_{\varphi \varphi}$ (usual $\alpha$-effect) in the dynamo process
 is well understood
as being directly related to the generation 
of a poloidal field from a toroidal field, the role of
the other coefficients of $\balpha$ is less clear and would need a specific
study in itself. 
\subsection{Numerical results}
\label{sec:results}
\begin{figure}
\begin{tabular}{@{}c@{\hspace{5mm}}c@{\hspace{2cm}}c@{}}
$\delta^{-1} \tilde{a}_{ss}$ &$\delta^{-3}\omega^{-1}\tilde{a}_{s\varphi}$&\\
    \epsfig{file=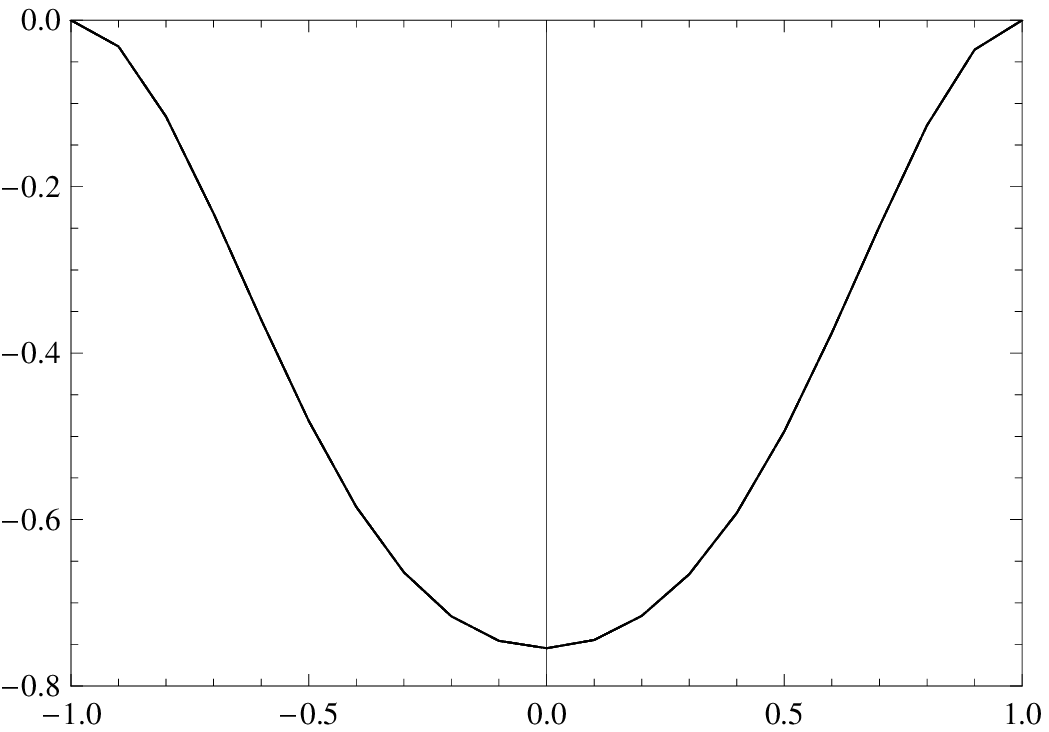,width=0.25\textwidth}
    &
    \epsfig{file=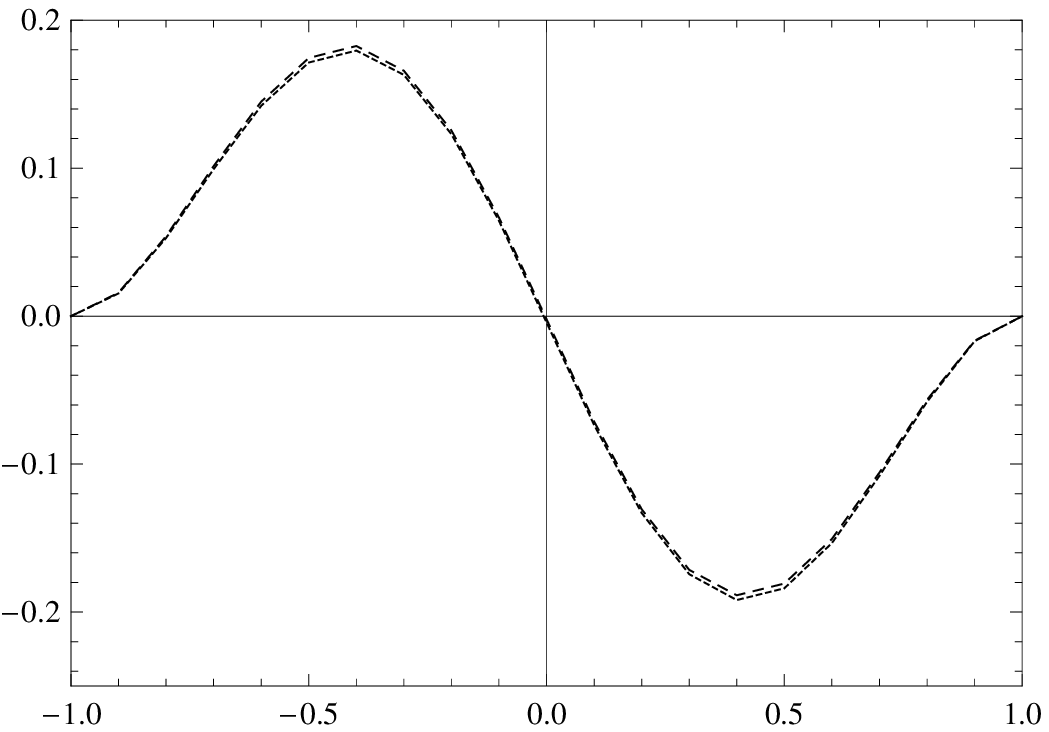,width=0.25\textwidth}
    &    
    \raisebox{1.7cm}{$\tilde{a}_{sz} = 0$}
    \\*[+0cm]
$\delta^{-3}\omega^{-1} \tilde{a}_{\varphi s}$ &$\delta^{-1}\tilde{a}_{\varphi\varphi}$&\\
    \epsfig{file=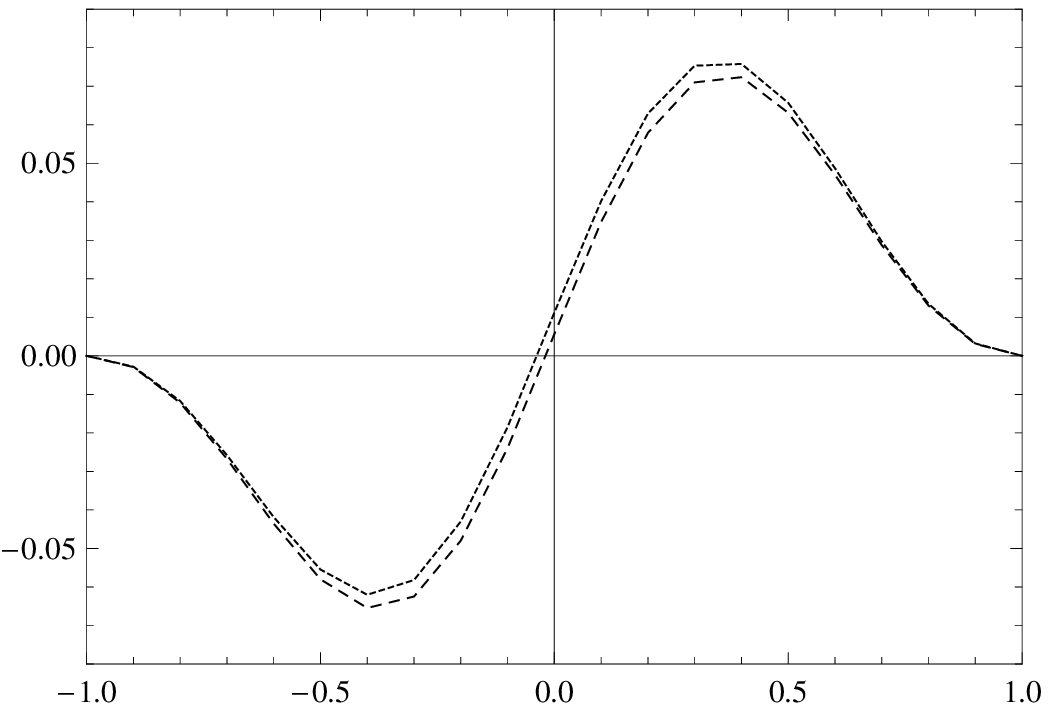,width=0.25\textwidth}
    &
    \epsfig{file=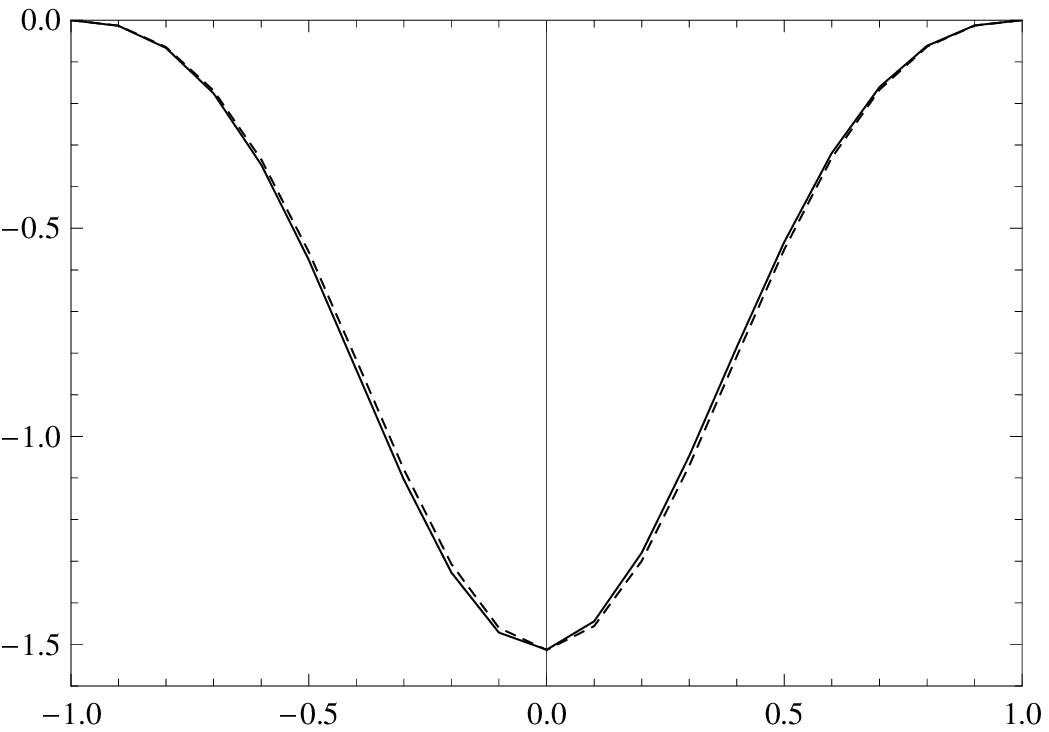,width=0.25\textwidth}
    &    
    \raisebox{1.7cm}{$\tilde{a}_{\varphi z} = 0$}
    \\*[+0cm]
$\delta^{-3}\omega^{-1} \tilde{a}_{zs}$ &$\delta^{-1}\tilde{a}_{z\varphi}$&\\
    \epsfig{file=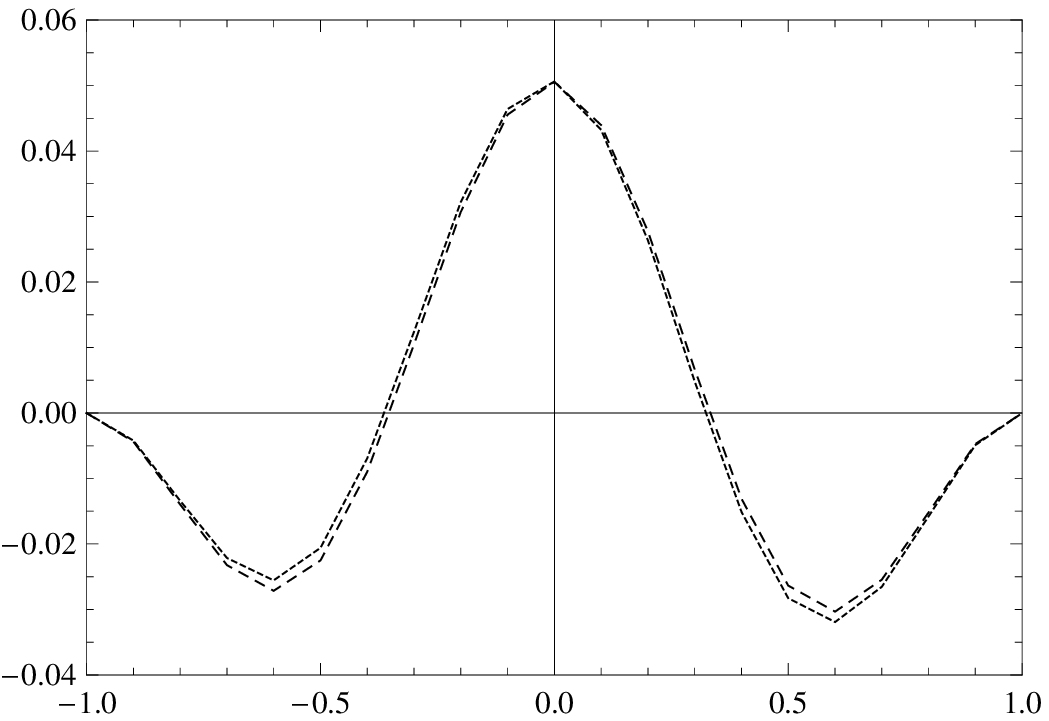,width=0.25\textwidth}
    &
    \epsfig{file=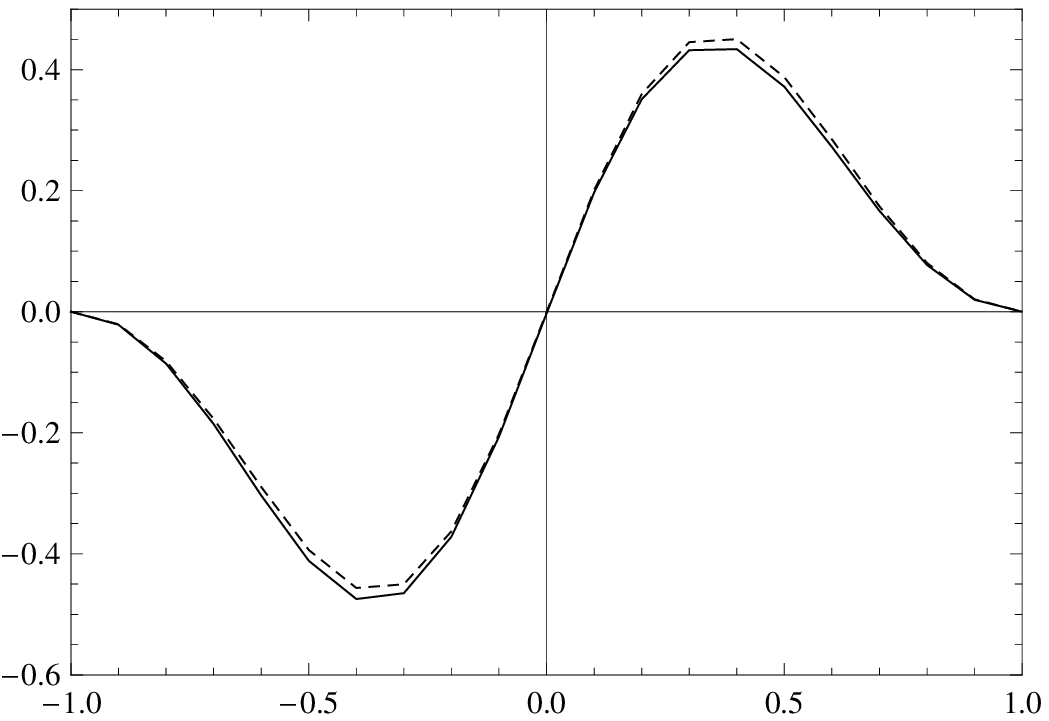,width=0.25\textwidth}
    &
    \raisebox{1.7cm}{$\tilde{a}_{zz} = 0$}
    \\*[+0cm]
  \end{tabular}
\caption{Scaling and $s$-profile of coefficients ${\tilde{a}}_{\kappa \lambda}$, 
for $\delta m = \pi / 2$, $\delta \ll 1$ and $\delta^2 \omega \ll 1$, and for rolls of type (ia).}
\label{fig:ia}
\end{figure}

\begin{figure}
\begin{tabular}{@{}c@{\hspace{5mm}}c@{\hspace{2cm}}c@{}}
$\delta^{-3} \omega^{-1} \tilde{a}_{ss}$ &$\delta^{-1}\tilde{a}_{s\varphi}$&\\
    \epsfig{file=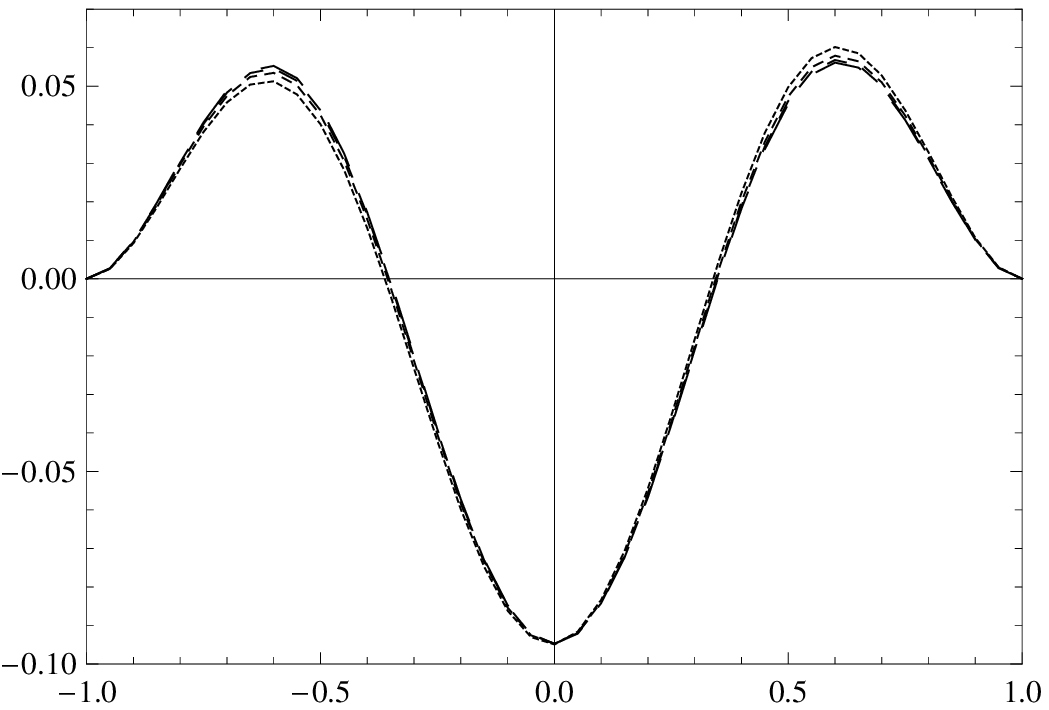,width=0.25\textwidth}
    &
    \epsfig{file=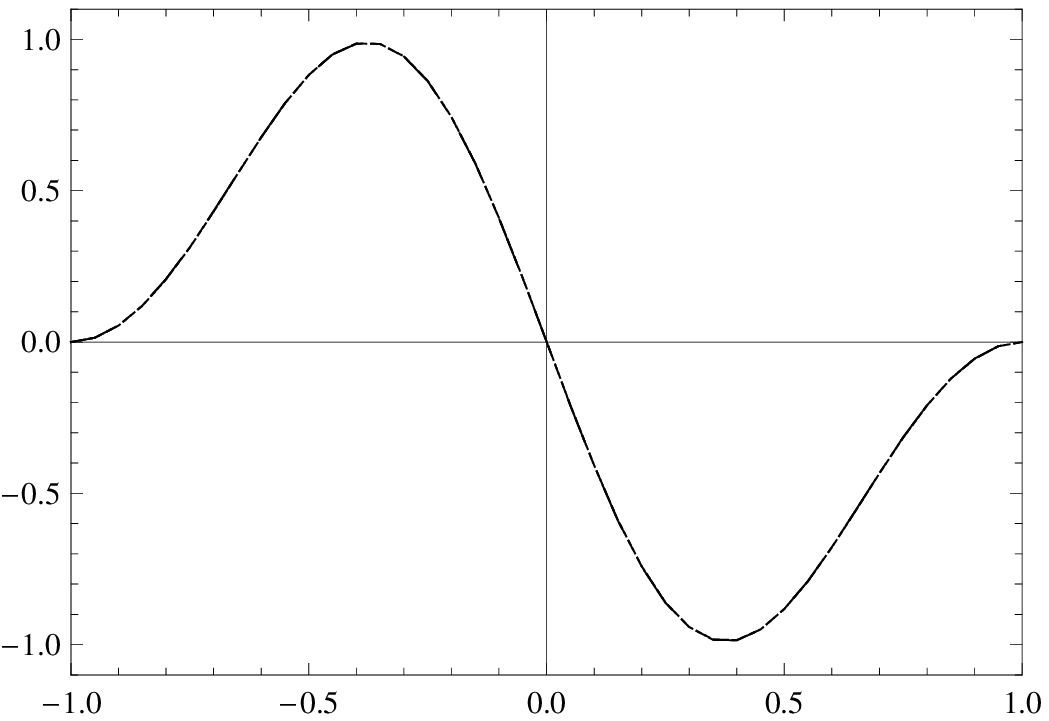,width=0.25\textwidth}
    &    
    \raisebox{1.7cm}{$\tilde{a}_{sz} = 0$}
    \\*[+0cm]
$\delta^{-1}\tilde{a}_{\varphi s}$ &$\delta^{-3}\omega^{-1}\tilde{a}_{\varphi\varphi}$&\\
    \epsfig{file=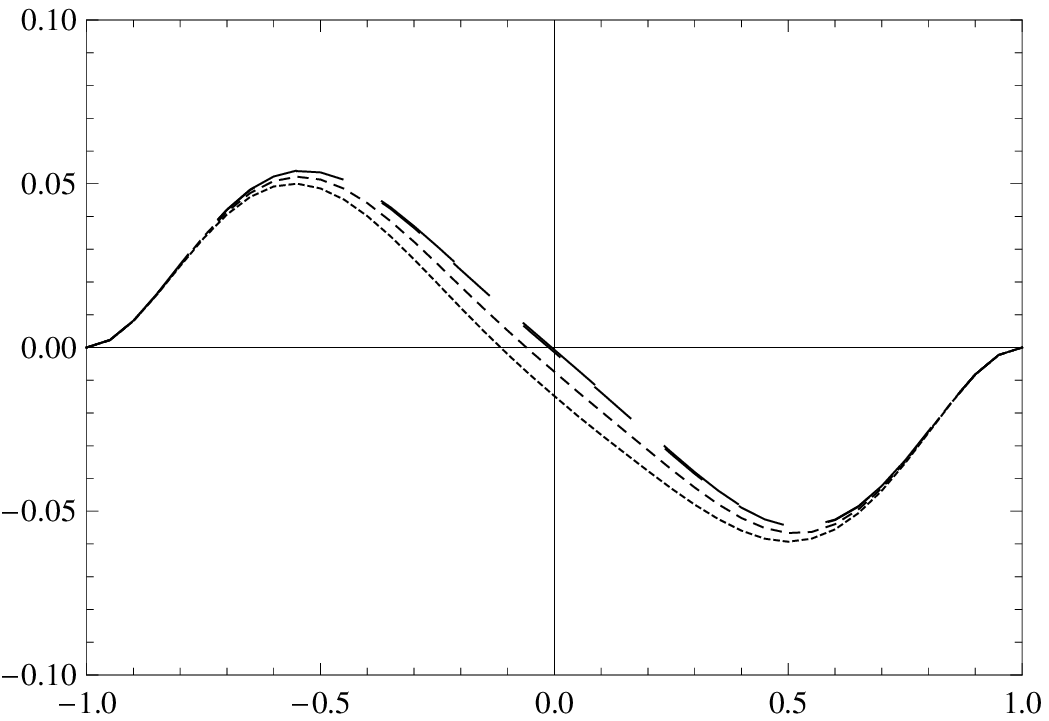,width=0.25\textwidth}
    &
    \epsfig{file=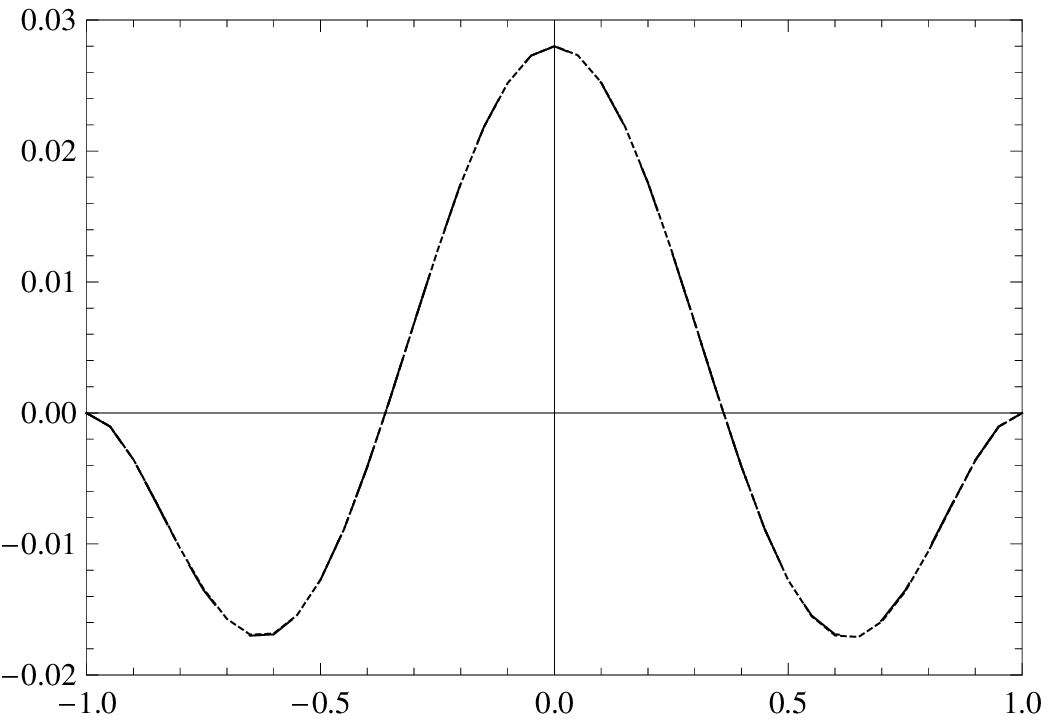,width=0.25\textwidth}
    &    
    \raisebox{1.7cm}{$\tilde{a}_{\varphi z} = 0$}
    \\*[+0cm]
$\delta^{-3}\omega^{-1} \tilde{a}_{zs}$ &$\delta^{-1}\tilde{a}_{z\varphi}$&\\
    \epsfig{file=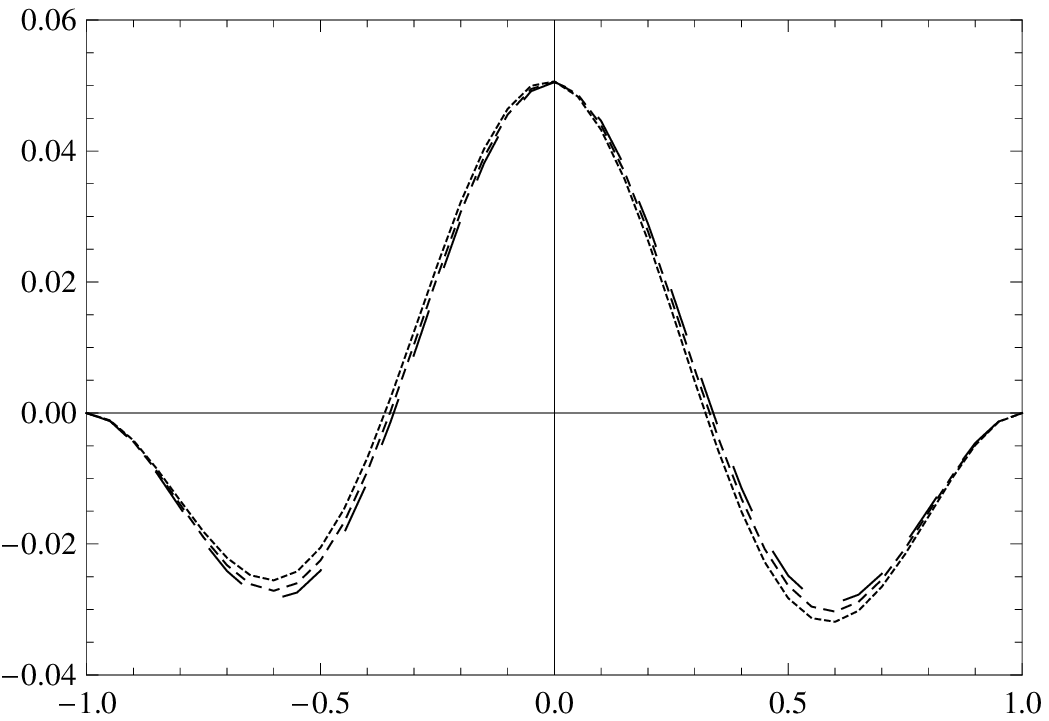,width=0.25\textwidth}
    &
    \epsfig{file=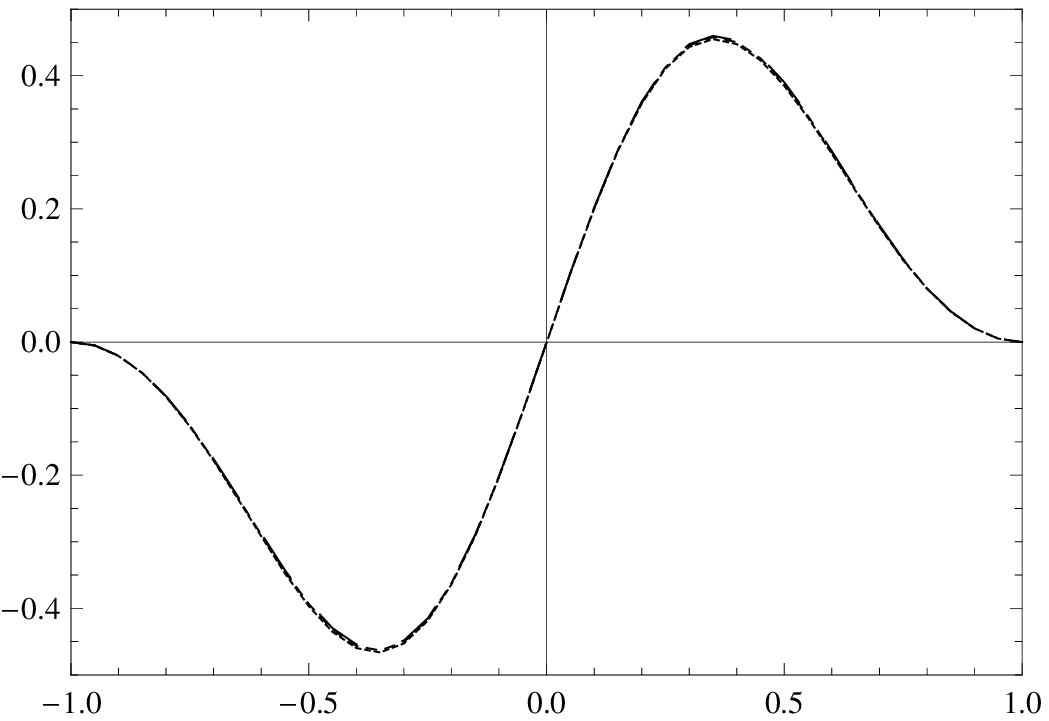,width=0.25\textwidth}
    &
    \raisebox{1.7cm}{$\tilde{a}_{zz} = 0$}
    \\*[+0cm]
  \end{tabular}
\caption{Same caption as figure \ref{fig:ia} but for rolls of type (iia).}
\label{fig:iia}
\end{figure}

\begin{figure}
\begin{tabular}{@{}c@{\hspace{5mm}}c@{\hspace{2cm}}c@{}}
$\delta^{-1} \tilde{a}_{ss}$ &$\delta^{-2}\tilde{a}_{s\varphi}$&\\
    \epsfig{file=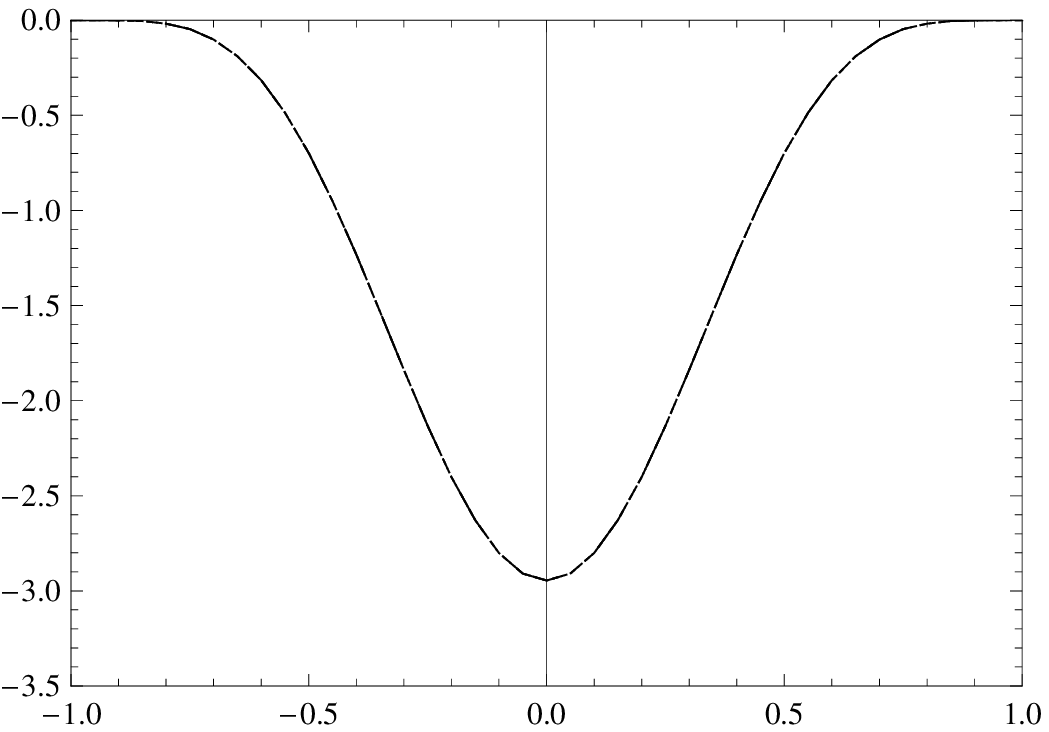,width=0.25\textwidth}
    &
    \epsfig{file=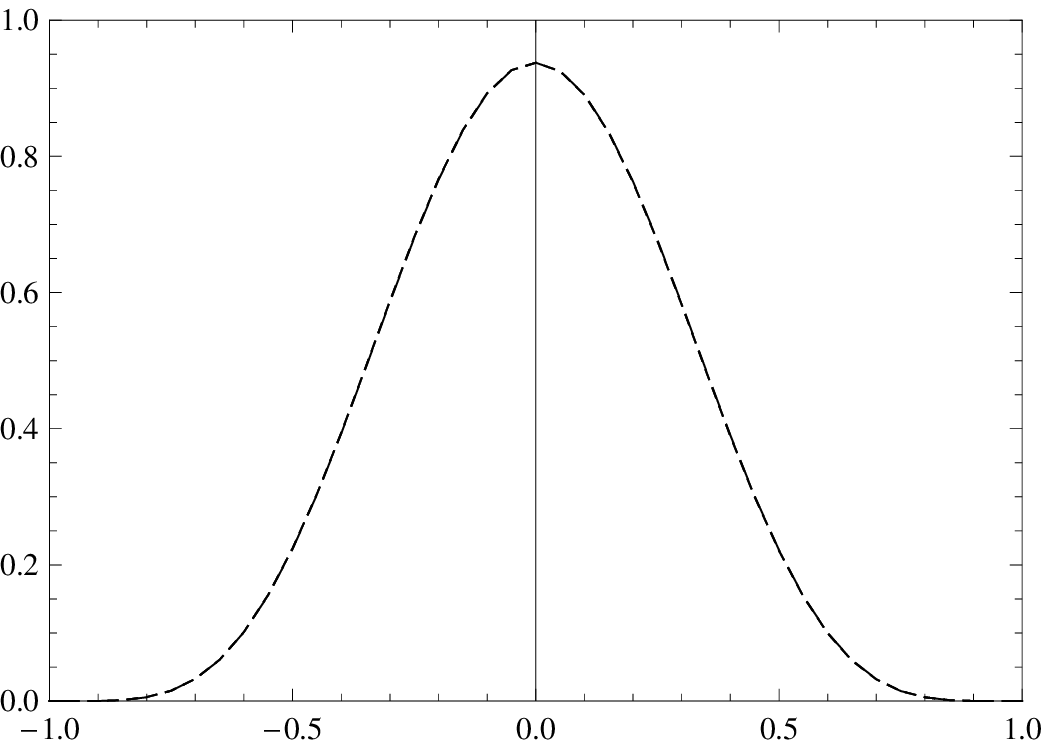,width=0.25\textwidth}
    &    
    \raisebox{1.7cm}{$\tilde{a}_{sz} = 0$}
    \\*[+0cm]
$\delta^{-2}\tilde{a}_{\varphi s}$ &$\delta^{-3}\tilde{a}_{\varphi\varphi}$&\\
    \epsfig{file=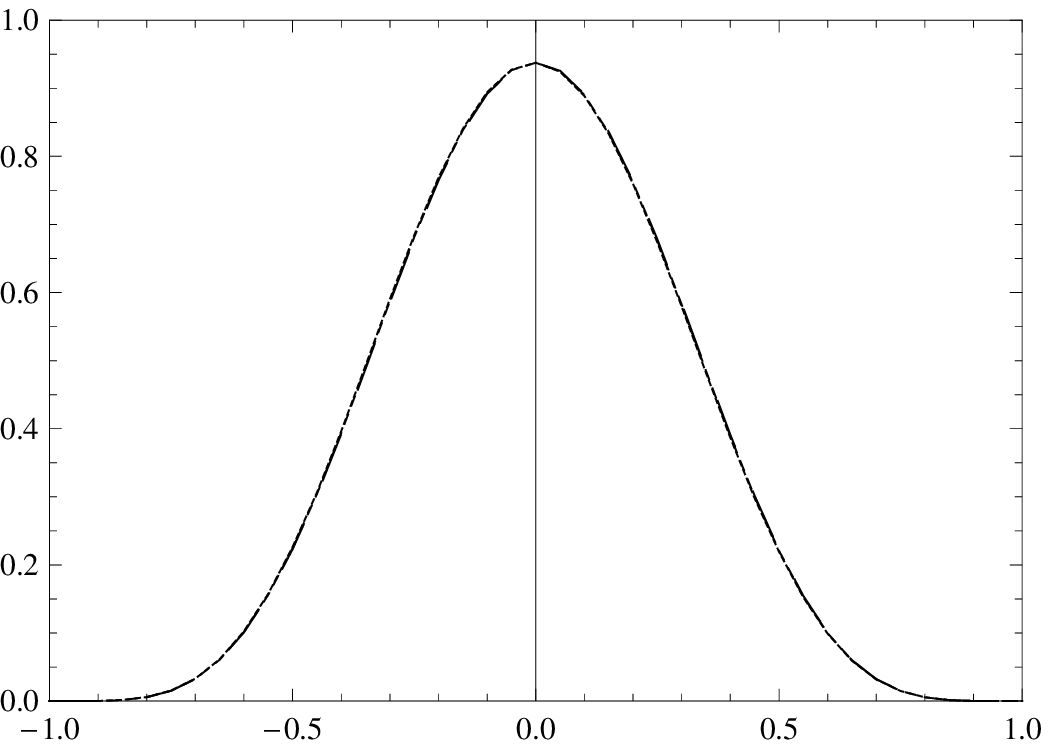,width=0.25\textwidth}
    &
    \epsfig{file=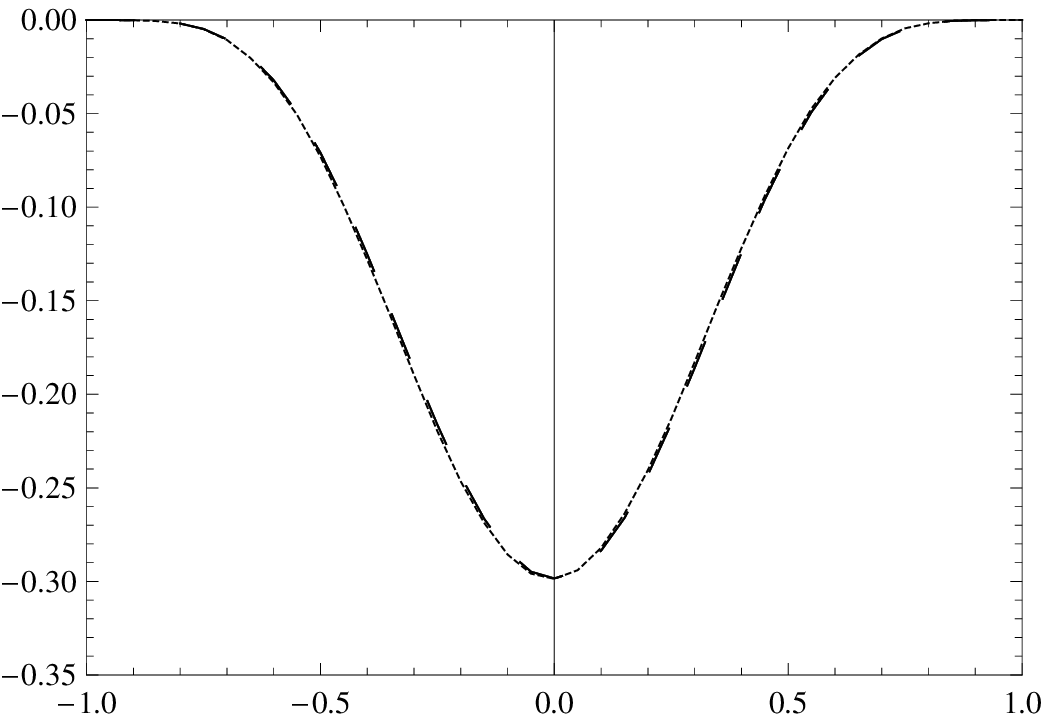,width=0.25\textwidth}
    &    
    \raisebox{1.7cm}{$\tilde{a}_{\varphi z} = 0$}
    \\*[+0cm]
$\delta^{-2} \tilde{a}_{zs}$ &$\delta^{-3}\tilde{a}_{z\varphi}$&\\
    \epsfig{file=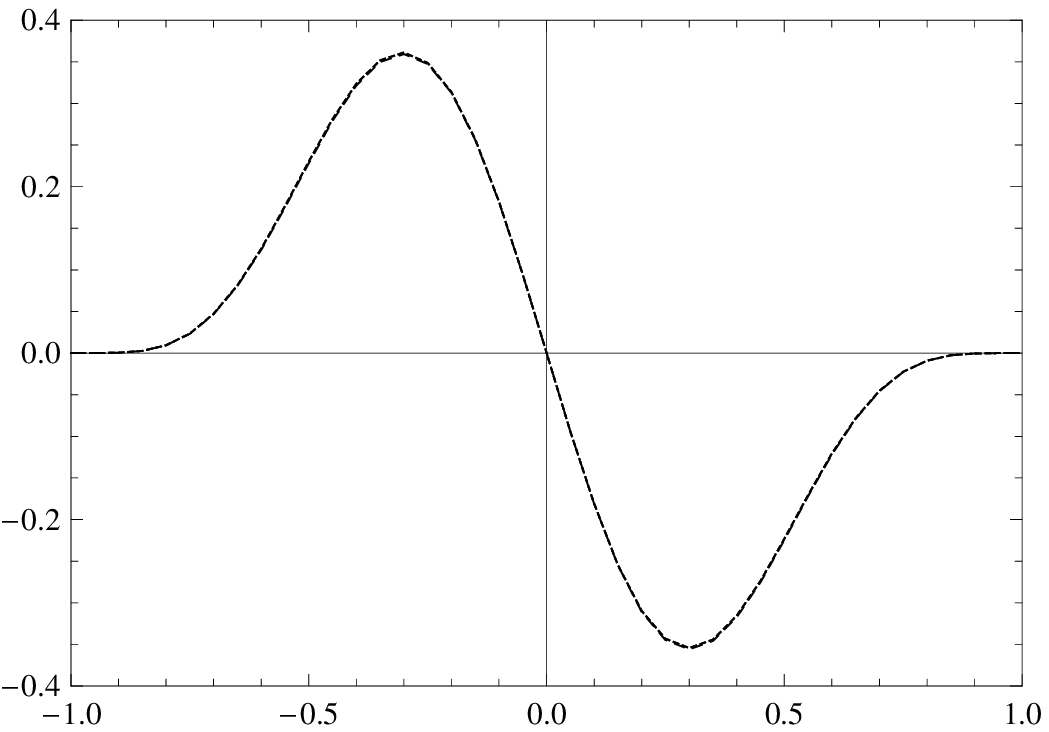,width=0.25\textwidth}
    &
    \epsfig{file=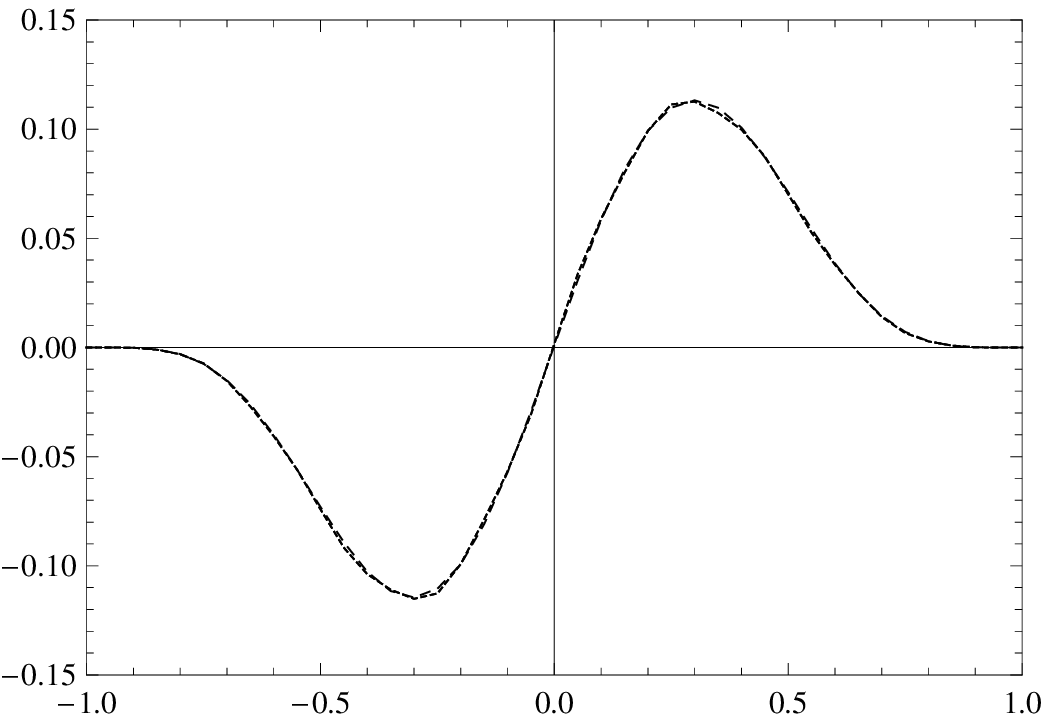,width=0.25\textwidth}
    &
    \raisebox{1.7cm}{$\tilde{a}_{zz} = 0$}
    \\*[+0cm]
  \end{tabular}
\caption{Same caption as figure \ref{fig:ia} but for rolls of type (ib).}
\label{fig:ib}
\end{figure}

\begin{figure}
\begin{tabular}{@{}c@{\hspace{5mm}}c@{\hspace{2cm}}c@{}}
$\delta^{-2} \tilde{a}_{ss}$ &$\delta^{-3}\tilde{a}_{s\varphi}$&\\
    \epsfig{file=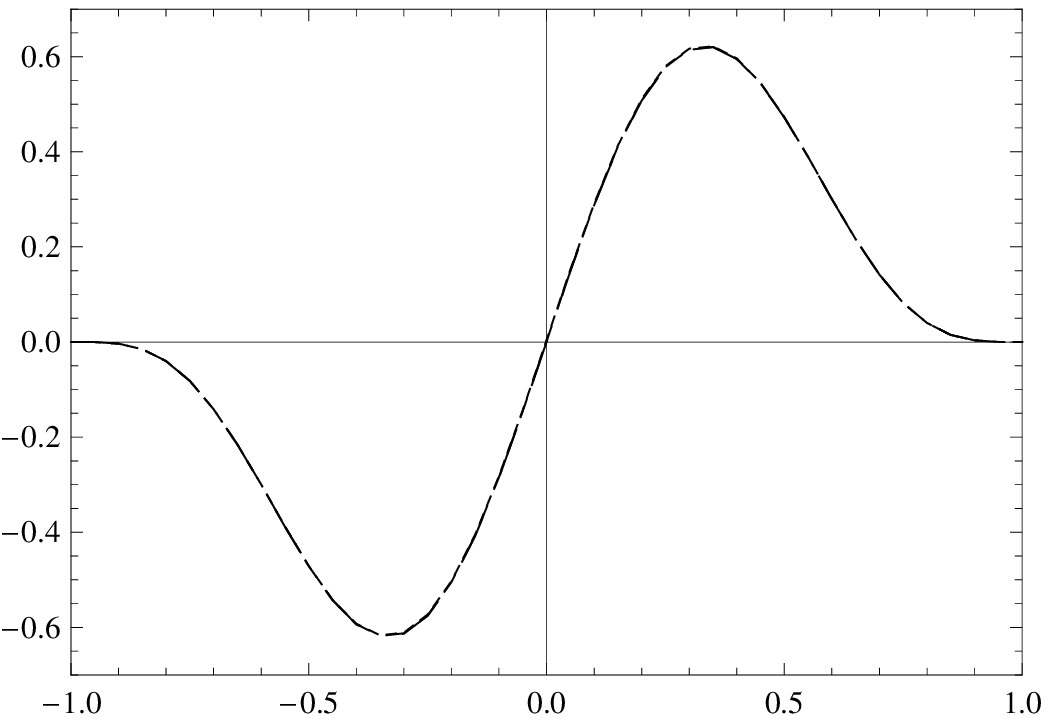,width=0.25\textwidth}
    &
    \epsfig{file=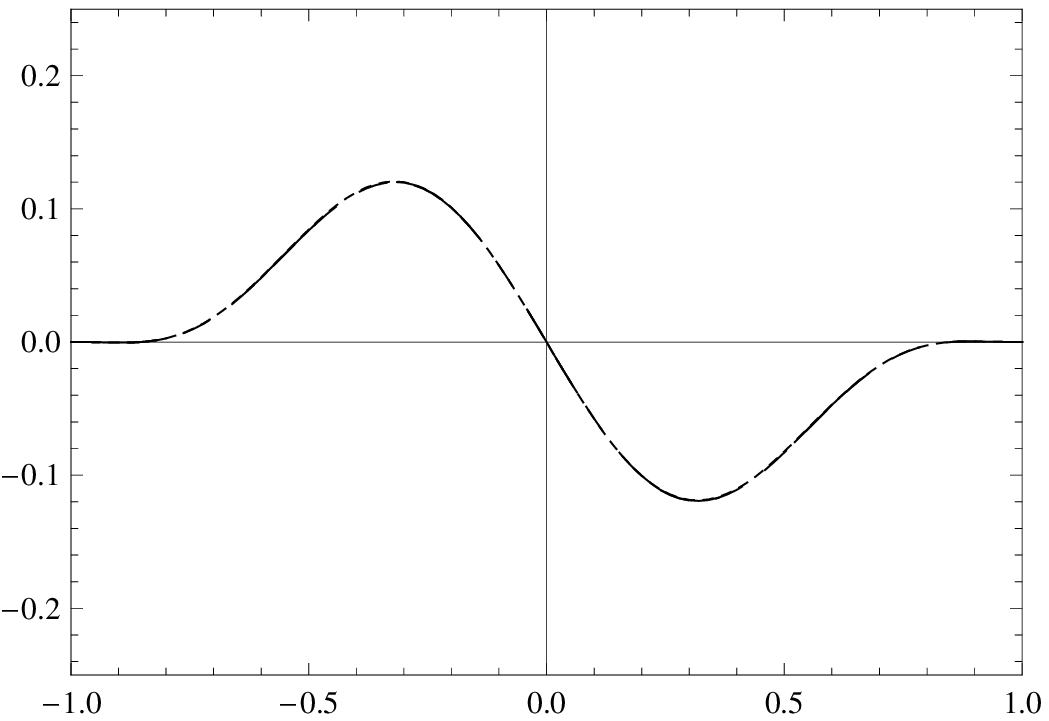,width=0.25\textwidth}
    &    
    \raisebox{1.7cm}{$\tilde{a}_{sz} = 0$}
    \\*[+0cm]
$\delta^{-3}\tilde{a}_{\varphi s}$ &&\\
    \epsfig{file=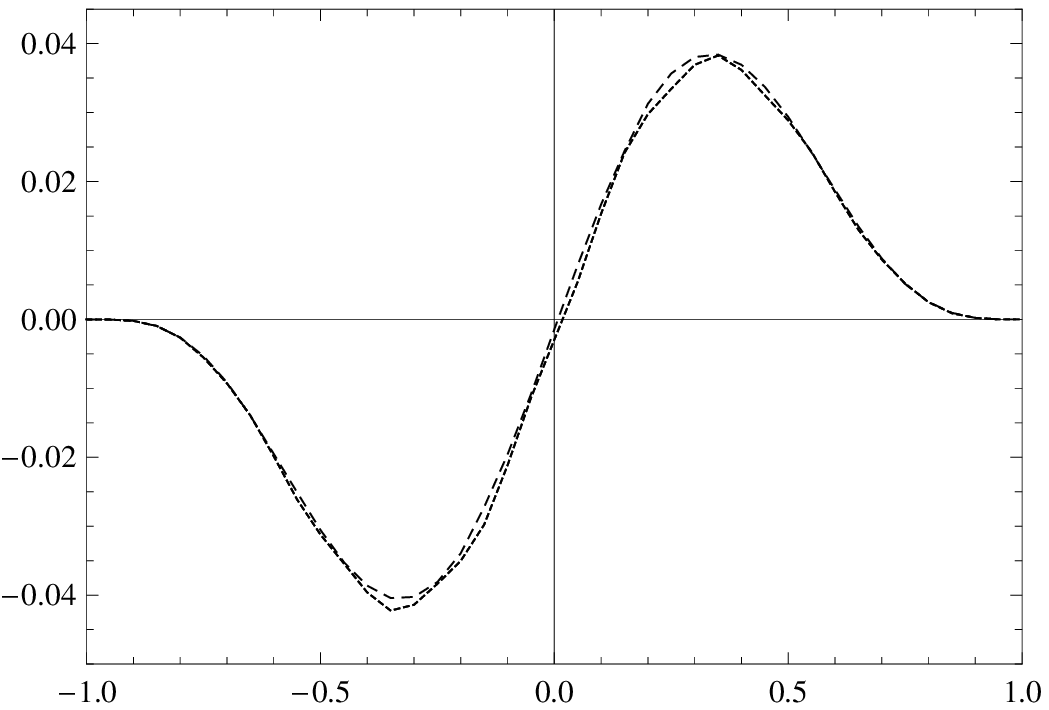,width=0.25\textwidth}
    &
    \raisebox{1.7cm}{$\tilde{a}_{\varphi\varphi} = 0$}
    &    
    \raisebox{1.7cm}{$\tilde{a}_{\varphi z} = 0$}
    \\*[+0cm]
$\delta^{-2} \tilde{a}_{zs}$ &$\delta^{-3}\tilde{a}_{z\varphi}$&\\
    \epsfig{file=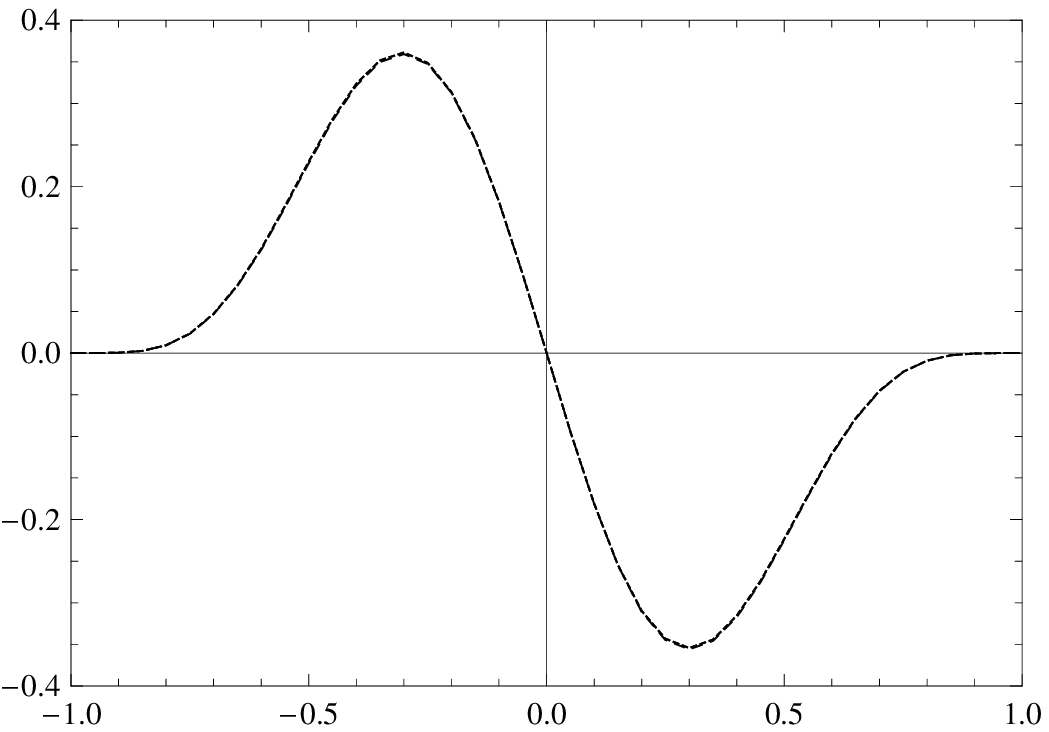,width=0.25\textwidth}
    &
    \epsfig{file=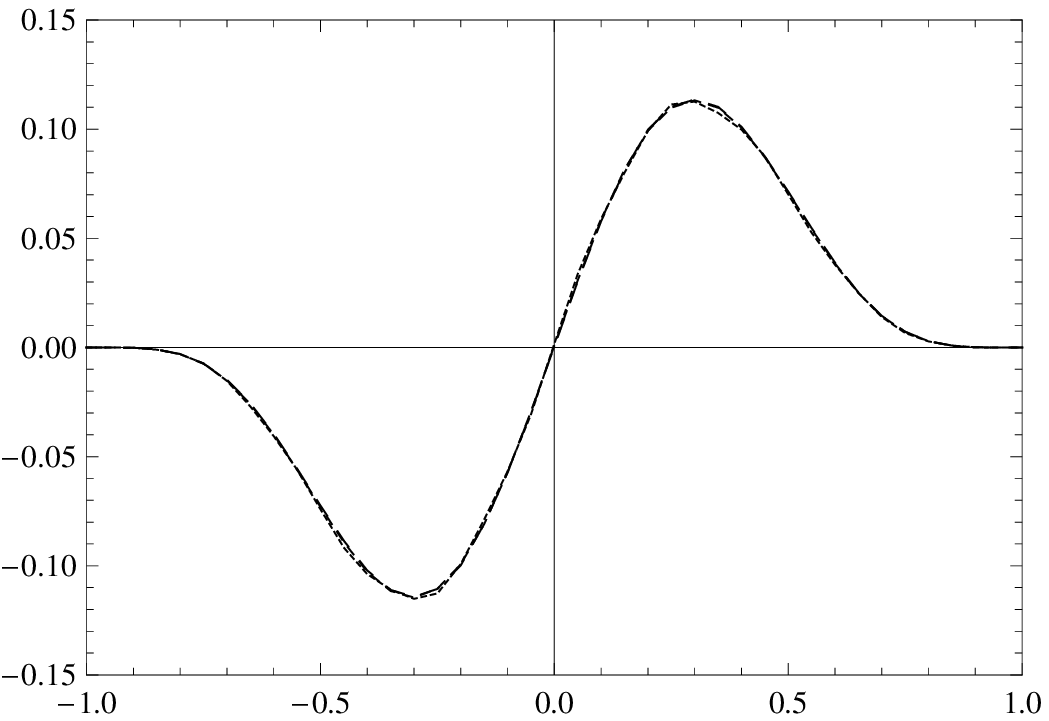,width=0.25\textwidth}
    &
    \raisebox{1.7cm}{$\tilde{a}_{zz} = 0$}
    \\*[+0cm]
  \end{tabular}
\caption{Same caption as figure \ref{fig:ia} but for rolls of type (iib).}
\label{fig:iib}
\end{figure}

In this section we plot the coefficients $\tilde{a}_{\kappa \lambda}$ for flows of type (ia) in figure \ref{fig:ia}, (iia) in figure \ref{fig:iia}, (ib) in figure \ref{fig:ib} and (iib) in figure \ref{fig:iib}.
We consider rolls satisfying $\delta m = \pi / 2$ and show results in the double asymptotic limit $\delta \ll 1$ and $\delta^2 \omega \ll 1$. We find that the profile of each coefficients $\tilde{a}_{\kappa \lambda}$ converge in this double limit. We also find some scaling laws in  $\delta$ and $\omega$ such that 
\begin{equation}
	\tilde{a}_{\varphi\varphi} \sim \omega^p \delta^q.
	\label{scaling}
\end{equation}
We found that their validity prevails even for $\omega \delta^2 \approx 1$.
For flows of type (ia) and (iia) the scalings 
in $\omega$ is consistent with the general structure of the kernel $K$ given in (\ref{kernel}) for $\omega=0$.

For flows of type (ib) and (iib) the scalings are rather different from cases (ia) and (iia). In particular we found no $\omega$ dependency,
suggesting that the leading term in $k_m(s,s')$ is of order $\delta$. Surprisingly enough in case (iib) we found that ${\tilde{a}}_{\varphi \varphi}\sim 0$ ruling out any chance to explain the dynamo mechanism with a simple $\alpha$-effect.

It is interesting to compare the results obtained for the $\alpha$-effect coefficient $\alpha_{\perp}$ in the case of the Roberts flow \citep{Radler02a} to our results for $a_{\varphi \varphi}$ for a flow of type (i) with compact cells. The length $a$ used there corresponds to $2 \delta l_0$, and the Reynolds numbers
$Rm_{\perp}$ and $Rm_{\parallel}$ used there have to be interpreted as $2 \delta R_{m \perp}$ and $\delta R_{m \parallel}$, respectively, with our $R_{m \perp}$ and $R_{m \parallel}$.
In that sense the result 
$\alpha_{\perp} = (\pi^2 / 16)(\eta / a)Rm_{\perp} Rm_{\parallel} \phi(Rm_{\perp}) $
reported in the mentioned paper 
takes the form
$\alpha_{\perp} = (\pi^2 / 16)(\eta \delta / l_0)R_{m\perp} R_{m\parallel} \phi(R_{m\perp}) $.
This quoted result applies to arbitrary $Rm_{\perp}$ and $ Rm_{\parallel}$. The function
$\phi$ is equal to unity for $Rm_{\perp}=0$ and so in the second-order correlation approximation.
It decreases monotonically if $Rm_{\perp}$ grows and tends to zero as $Rm_{\perp} \rightarrow \infty$.
According to (\ref{eq:abHZ}) we have $a_{\varphi \varphi} = (\eta / l_0)R_{m \perp} R_{m \parallel} \tilde{a}_{\varphi \varphi}$.
The same relation applies with the averages $<a_{\varphi \varphi}>$ and $<\tilde{a}_{\varphi \varphi}>$
of $a_{\varphi \varphi}$ and $\tilde{a}_{\varphi \varphi}$ over $s$. We take from figure 
\ref{fig:ia} that $<\tilde{a}_{\varphi \varphi}> \approx \delta$. This leads to 
$<a_{\varphi \varphi}> \approx (\eta \delta / l_0)R_{m\perp} R_{m\parallel} $.
Hence our result for $a_{\varphi \varphi}$ (derived in the second-order correlation approximation)
is in reasonable agreement with the result for the Roberts flow.\\
The comparison with the result for the Roberts flow suggest that our result for $a_{\varphi \varphi}$
remains valid for all values of $R_{m \parallel}$ and, as can be concluded from the specific properties of $\phi$, for values of $\delta R_{m \parallel}$ up to the order of unity. 
There is, however, no straightforward extension of the proof for the linearity in $R_{m\parallel}$
to our case.
We further learn here that it is of less importance for the magnitude of the $\alpha$-effect whether or not a given roll is at all sides surrounded by other rolls.

\section{Conclusions}
\label{sec:Discussion}
The main insight of this study is that the $\alpha$-effect and presumably the dynamo action generated by Rossby waves depends drastically on the Ekman number $E$. The reason for that lies in the competition between
Ekman pumping and geometrical slope effect as the main source of the axial flow, producing different phase shifts between the horizonthal and vertical components of the flow. In the limit $E\ll 1$ the geometrical slope effect prevails.
In particular we find that the mean-field coefficients have then different radial profiles and, probably more important, different asymptotic scalings in the limit of small rolls radius $\delta\ll 1$ and wave frequency such that $\delta^2 \omega\ll1$.

In addition we found that for rolls with an arbitrary radial shift the $\alpha$-tensor is completely changed not only in terms of the radial profile of the tensor coefficients but also in terms of scalings in $\delta$ and $\omega$.
In particular we found that the $\alpha$-effect corresponding to the coefficient ${\tilde{a}}_{\varphi \varphi}$
may disappear, making then difficult the interpretation of the dynamo mechanism if any.

There are two recent numerical studies in which the $\alpha$--tensor produced
by Rossby waves has been calculated.
\cite{Schrinner05, Schrinner06} extracted the $\alpha$--tensor (and higher-order coefficients) from geodynamo simulations
with $E = 10^{-3}$.
\cite{Schaeffer06} did the same with simulations
of the Taylor--Couette convection using the quasi--geostrophic approximation
with values of $E$ down to $10^{-8}$.
It is then of interest to compare qualitatively our results to these findings (ignoring the difference addressed at the end of section \ref{sub:coeff}
between the two definitions of the $\alpha$--effect
$a_{\varphi \varphi}$ and $\breve{a}_{\varphi \varphi}$).

In the study by \cite{Schrinner05, Schrinner06} a thermally driven dynamo in a rotating spherical shell
is considered.
In a regime not too far from to the onset of convection, drifting rolls are observed.
As a consequence of $E = 10^{-3}$,
the main cause of the axial flows in these rolls should be Ekman pumping.
Indeed the dependence of $\alpha_{\varphi \varphi}$ on $s$ corresponds roughly to that
of our ${\tilde{a}}_{\varphi \varphi}$ depicted in Figures~\ref{fig:ia}.
This shows that for not a too small Ekman number, the dynamo process can be understood on the basis of the mechanism described in figure \ref{heur}(i). 
 
\cite{Schaeffer06} dealt with the Rossby wave instabilities of a geostrophic internal shear layer produced by differential rotation between two spheres. They used the 
quasi-geostrophic approximation, including both Ekman pumping and slope effect as sources of the axial flow.
By inserting the corresponding velocity field in an induction equation solver they showed that it is capable of dynamo action. 
They also derived the $\alpha$ coefficients from their numerical simulation for $E=10^{-8}$. 
The dependence of their $\alpha_{\varphi \varphi}$ on $s$ corresponds roughly to that of our $\tilde{a}_{\varphi \varphi}$ when 
adapting our flow definition (\ref{eq227b}) to
the quasi-geostrophic approximation. The comparison of the other coefficients is left for future work.

Let us add a remark concerning the sign changes of ${\tilde{a}}_{\varphi \varphi}$
in the case (iia).
Investigating a spherical $\alpha$--effect dynamo model with isotropic $\alpha$--effect
and a spherically symmetric coefficient $\alpha$,  \cite{Stefani05}
found polarity reversals of the mean magnetic field if $\alpha$ changes its sign
along the radius (see also \citealt{Giesecke05a,Giesecke05b}).

In our calculations both the fluid flow and the magnetic field have been considered as independent of $z$.
Therefore the results reflect by far not all essential features of the convection rolls in a rotating liquid sphere.
A more realistic treatment of the problem would require to consider not only the $z$-dependencies mentioned
but also the multi-scales structure of the flow.
The difficulties that arise in this way could perhaps be reduced by taking advantage
of the quasi--geostrophic approximation.
It is then possible that only rolls of a certain scale are important for the dynamo process, that is, those having a sufficiently large $\delta R_m$ to produce dynamo action but not too large in order to avoid too strong flux expulsion and $\alpha$-quenching. In this case the simple picture of a ring of drifting rolls generating an $\alpha$-effect as one part of the dynamo mechanism (completed by differential rotation) might be still relevant.

\section*{Acknowledgments}
This study was supported by a grant from the LEGI (Appel d'offres LEGI 2002).
R.A.Z. was also supported by a Conacyt grant from Mexico.
FP and KHR are grateful to the Dynamo Program at the Kalvi Intitute for Theoretical Physics, Santa Barbara, California (supported in part by the National Science Foundation under Grant No. PHY05-51164) for completion of the paper. 
We thank N. Schaeffer, P. Cardin, D. Jault and M. Schrinner for interesting discussions.
We also thank the referee K. Zhang for suggesting the implementation of the radial phase-shift.

\appendices
\section{\vspace{12pt}\\General solutions of equations (\ref{eq203}) and (\ref{eq165}) in terms of Green's functions}
\label{mathematics}
In view of (\ref{eq203})
we consider first
the equation
\begin{equation}
(D_m - {\rm i} \omega) f(s) = g(s)
\label{eq171}
\end{equation}
in $0 \leq s < \infty$.
Its general solution reads
\begin{eqnarray}
f (s) &=& c_1 I_m (\sqrt{{\rm i}\omega}\; s) + c_2 K_m (\sqrt{{\rm i}\omega} s)
\nonumber\\
&&  - K_m (\sqrt{{\rm i}\omega}\;s) \int_0^s I_m (\sqrt{{\rm i}\omega}\;s') \, g (s',k) \, s' \, \dd s'\label{eq259} \\
&&  - I_m (\sqrt{{\rm i}\omega}\;s) \int_s^\infty K_m (\sqrt{{\rm i}\omega}\;s') \, g (s',k) \, s' \, \dd s' \, ,
\nonumber
\end{eqnarray}
where $c_1$ and $c_2$ are arbitrary constants,
and $I_m$ and $K_m$ modified Bessel functions of the first and second kind
(see Appendix \ref{sec:Bessel}).
In order to exclude
singularities at $s=0$ and for $s \to \infty$,  we
use (\ref{eq421}) and (\ref{eq423}),
 put $c_1 = c_2 = 0$
and obtain
\begin{equation}
f (s) = - \int_0^\infty k_m (s,s') \, g (s') \, s' \, \dd s' \, ,
\label{eq260}
\end{equation}
with a Green's function $k_m$ defined by
\begin{eqnarray}
k_m (s, s') =  K_m (\sqrt{{\rm i}\omega}\; s) I_m (\sqrt{{\rm i}\omega}\; s') & \mbox{for} & s' \leq s
\nonumber\\
k_m (s, s') =  I_m (\sqrt{{\rm i}\omega}\; s) K_m (\sqrt{{\rm i}\omega}\; s') & \mbox{for} & s \leq s' \, .
\label{eq261append2}
\end{eqnarray}
Applying this to (\ref{eq203}),
we find
\begin{equation}
\hat{T}(s) = \int_0^\infty k_m (s,s') \, \hat{Q}_z (s') \, s' \, \dd s', \quad \quad
\hat{S}(s) = - \int_0^\infty k_m (s,s') \, \hat{F} (s') \, s' \, \dd s'.
\label{eq2602}
\end{equation}
The Green's function $k_m$ defined by (\ref{eq261append2})
is complex. Its 
real and imaginary parts $k_m^R$ and $k_m^I$ can be written in the form:
\begin{eqnarray}
	k_m^R(s,s') &=& \ker_m(\sqrt{\omega}s)\; \ber_m(\sqrt{\omega}s') - \kei_m(\sqrt{\omega}s)\; \bei_m(\sqrt{\omega}s') \quad \mbox{for} \quad s' \le s \nonumber \\
		&=& \ber_m(\sqrt{\omega}s)\; \ker_m(\sqrt{\omega}s') - \bei_m(\sqrt{\omega}s)\; \kei_m(\sqrt{\omega}s') \quad \mbox{for} \quad s' \ge s \label{kmR}\\
	k_m^I(s,s') &=& \kei_m(\sqrt{\omega}s)\; \ber_m(\sqrt{\omega}s') + \ker_m(\sqrt{\omega}s)\; \bei_m(\sqrt{\omega}s') \quad \mbox{for} \quad s' \le s \nonumber \\
		&=& \bei_m(\sqrt{\omega}s)\; \ker_m(\sqrt{\omega}s') + \ber_m(\sqrt{\omega}s)\; \kei_m(\sqrt{\omega}s') \quad \mbox{for} \quad s' \ge s	\label{kmI}
\end{eqnarray}
where $\ber_m$, $\bei_m$, $\ker_m$ and $\kei_m$ are the Kelvin functions of order $m$.\\

In view of (\ref{eq165}) we consider first
the general ordinary differential equation
\begin{equation}
D_m f(s) = g(s)
\label{eq1713}
\end{equation}
in $0 \leq s < \infty$.
Its general solution reads
\begin{equation}
f (s) = c_1 s^m + c_2 s^{-m}
   - \frac{1}{2 m} \big(s^{-m} \int_0^s g (s') \, {s'}^{m+1} \, \dd s'
   + s^m \int_s^\infty g (s') \, {s'}^{-m+1} \, \dd s' \big) \, ,
\label{eq1733}
\end{equation}
where $c_1$ and $c_2$ are again arbitrary constants.
In order to exclude singularities at $s=0$ and for $s \to \infty$ we put $c_1 = c_2 = 0$.
In this way we obtain
\begin{equation}
f (s) = - \int_0^\infty h_m (s, s') \, g (s') \, s' \, \dd s'
\label{eq1743}
\end{equation}
with a Green's function $h_m$ 
\begin{eqnarray}
h_m (s, s') = \frac{1}{2m} \big(\frac{s'}{s}\big)^m & \mbox{for} & s' \leq s
\nonumber\\
h_m (s, s') = \frac{1}{2m} \big(\frac{s}{s'}\big)^m & \mbox{for} & s \leq s' \, .
\label{eq1753}
\end{eqnarray}
As to be expected, $k_m$ turns into $h_m$ if $\omega = 0$.
\section{\vspace{12pt}\\Useful relations for modified Bessel functions}
\label{sec:Bessel}
The modified Bessel functions of first kind, $I_m (x)$, and of second kind, $K_m (x)$,
are solutions of the equation
\begin{equation}
x^2 y'' + x y' -(x^2 + m^2) y = 0, \;\;\; m \ge 0.\nonumber
\end{equation}
From the recurrence relations
\begin{eqnarray}
	I'_m(x) &=& +I_{m-1}(x) - \frac{m}{x}I_m(x) \nonumber \\
	K'_m(x) &=& -K_{m-1}(x) - \frac{m}{x}K_m(x)\nonumber
\end{eqnarray}
and the Wronskian relation
\begin{equation}
I_m (x) K_{m-1} (x) + I_{m-1} (x) K_m (x)
    = I_m (x) K_{m+1} (x) + I_{m+1} (x) K_m (x)  = \frac{1}{x} \nonumber
\end{equation}
we conclude that
\begin{equation}
I_m (x) K_m^\prime (x) - I_m^\prime (x) K_m (x)  = - \frac{1}{x} \, .
\label{eq413}
\end{equation}
The modified Bessel functions satisfy
the integral relations
\begin{eqnarray}
\int x^{m+1} I_m (x) \dd x &=& x^{m+1} I_{m+1} (x) \, , \quad
   \int x^{-m+1} I_m (x) \dd x = x^{-m+1} I_{m-1} (x)
\nonumber\\
\int x^{m+1} K_m (x) \dd x &=& - x^{m+1} K_{m+1} (x) \, , \quad
   \int x^{-m+1} K_m (x) \dd x = - x^{-m+1} K_{m-1} (x) \nonumber
\end{eqnarray}
as well as the asymptotic relations
\begin{eqnarray}
I_m (x) &\approx& \frac{1}{m!} \, (\frac{x}{2})^m \,
   \big(1 + \frac{1}{m+1}(\frac{x}{2})^2 \big)
   \quad \mbox{as} \quad x \to 0
\nonumber\\
K_1 (x) &\approx& \frac{1}{x} \,
   \big(1 + \frac{x^2}{2} \, \log (\frac{x}{2}) \big)
   \quad \mbox{as} \quad x \to 0
\label{eq421}
\\
K_m (x) &\approx& \frac{(m-1)!}{2} \, (\frac{x}{2})^{-m} \,
   \big(1 - \frac{1}{m-1}(\frac{x}{2})^2 \big)
   \quad \mbox{as} \quad x \to 0
   \quad \mbox{if} \quad m \geq 2 \, .
\nonumber
\end{eqnarray}
and
\begin{equation}
I_m (x) \approx \frac{1}{\sqrt{2 \pi x}} \, {\rm e}^x \, , \quad
    K_m (x) \approx \sqrt{\frac{\pi}{2 x}} \, {\rm e}^{-x}  \quad \mbox{as}\quad x \to \infty.
\label{eq423}
\end{equation}


\begin{thebibliography}{9}
%
%
     \bibitem[\protect\citeauthoryear{Aubert \textit{et al.}}{2001}]{Aubert01}
     Aubert, J., Brito, D., Nataf, H.-C., Cardin, P. and Masson, J.-P.,
     A systematic experimental study of spherical shell convection in water and liquid gallium,
     \textit{Phys. Earth Planet. Inter.,} 2001 \textbf{128}, 51-74.

     \bibitem[\protect\citeauthoryear{Aubert \textit{et al.}}{2003}]{Aubert03}
     Aubert, J., Gillet, N. and Cardin, P.,
     Quasigeostrophic models of convection in rotating spherical shells,
     \textit{G-cubed,} 2003 \textbf{4}, 1052-1070.
     
     \bibitem[\protect\citeauthoryear{Braginskii} {1964a,b}]{Braginskii64a}
     Braginskii, S.I.,
     Self-excitation of a magnetic field during the motion of a highly conducting fluid,
     \textit{Sov. Phys. JETP,} 1964 \textbf{20}, 726-35.

     \bibitem[\protect\citeauthoryear{Braginskii} {1964}]{Braginskii64b}
     Braginskii, S.I.
     Theory of the hydromagnetic dynamo,
     \textit{Sov. Phys. JETP,} 1964 \textbf{20}, 1462-71.

     \bibitem[\protect\citeauthoryear{Busse}{1968}]{Busse68}
     Busse, F.H.
     Shear flow instabilities in rotating systems,
     \textit{J.~Fluid Mech.,} 1968 \textbf{33}, 557-589.

     \bibitem[\protect\citeauthoryear{Busse}{1970}]{Busse70}
     Busse, F.H.
     Thermal instabilities in rapidly rotating systems,
     \textit{J.~Fluid Mech.,} 1970 \textbf{44}, 441-460.
     
     \bibitem[\protect\citeauthoryear{Busse}{1975}]{Busse75}
     Busse, F.H. 
     A model of the Geodynamo,
     \textit{Geophys. J. R. Astron. Soc.,} 1975 \textbf{42}, 437-459.
     
     \bibitem[\protect\citeauthoryear{Busse}{1976}]{Busse76a}
     Busse, F.H. 
     Generation of planetary magnetism by convection,
     \textit{Phys. Earth Planet. Int.,} 1976 \textbf{42}, 437-459.
          
     \bibitem[\protect\citeauthoryear{Busse and Carrigan}{1976}]{Busse76b}
     Busse, F.H. and Carrigan, C.R. 
     Laboratory simulation of thermal convection in rotating planets and stars,
     \textit{Science,} 1976 \textbf{191},81.
     
     \bibitem[\protect\citeauthoryear{Busse}{2002}]{Busse02}
     Busse, F.H. 
     Convective flows in rapidly rotating spheres and their dynamo action,
     \textit{Phys.~ Fluid,} 2002 \textbf{14}, 1301-1314.

     \bibitem[\protect\citeauthoryear{Busse \textit{et al.}}{2005}]{Busse05}
     Busse, F.H., Zhang, K., Liao, X. 
     On slow inertial waves in the solar convection zone,
     \textit{The Astrophys. J.,} 2005 \textbf{631},171-174.

     \bibitem[\protect\citeauthoryear{Cardin and Olson}{1994}]{Cardin94}
     Cardin, P. and Olson, P. 
     Chaotic thermal convection in a rapidly rotating spherical shell: Consequences for flow in
     the outer core,
     \textit{Phys. Earth Planet. Inter.,} 1994 \textbf{82}, 235.

     \bibitem[\protect\citeauthoryear{Cardin and Olson}{1995}]{Cardin95}
     Cardin, P. and Olson, P. 
     The influence of the toroidal field on thermal convection in the core,
     \textit{Earth and Planet. Sc. Let.,} 1995 \textbf{132}, 167-181.

     \bibitem[\protect\citeauthoryear{Cardin \textit{et al.}}{2002}]{Cardin02}
     Cardin, P., Jault, D., Nataf, H.-C., Masson, J.-P. and Brito, D. 
     Towards a rapidly rotating liquid sodium dynamo experiment,
     \textit{Magnetohydrodynamics,} 2002 \textbf{38}, 177-189.

     \bibitem[\protect\citeauthoryear{Carrigan and Busse}{1983}]{Carrigan83}
     Carrigan, C.R. and Busse, F.H.
     An experimental and theoritical investigation ofthe onset of convection in rotating spherical
     shells,
     \textit{J.~Fluid Mech.,} 1983 \textbf{126},287.

     \bibitem[\protect\citeauthoryear{Dormy \textit{et al.}}{2004}]{Dormy04}
     Dormy, E., Soward, A.M., Jones, C.A., Jault, D. and Cardin, P. 
     The onset of thermal convection in a rotating spherical shells,
     \textit{J.~Fluid Mech.,} 2004 \textbf{501}, 43-70.
     
     \bibitem[\protect\citeauthoryear{Elsasser}{1956}]{Elsasser56}
     Elsasser, W.M. 
     Hydromagnetic Dynamo Theory,
     \textit{Rev. Modern Phys.,} 1956 \textbf{28}, 135-163.

 	   \bibitem[\protect\citeauthoryear{Giesecke \textit{et al.}}{2005a}]{Giesecke05a}
     Giesecke, A., Ziegler, U. and  R{\"u}diger, G. 
     Geodynamo alpha-effect derived from box simulations of rotating magnetoconvection,
     \textit{Phys. Earth Planet. Inter.,} 2005 \textbf{152}, 90-102.
     
 	   \bibitem[\protect\citeauthoryear{Giesecke \textit{et al.}}{2005b}]{Giesecke05b}
     Giesecke, A., R{\"u}diger, G. and Elstner, D. 
     Oscillating $\alpha^2$-dynamos and the reversal phenomenon of the global geodynamo,
     \textit{Astron. Nachr.,} 2005 \textbf{326}, 693-700.
     
     \bibitem[\protect\citeauthoryear{Grote and Busse}{2001}]{Grote01}
     Grote, E. and Busse, F.H. 
     Dynamics of convection and dynamos in rotating spherical fluid shells,
     \textit{Fluid Dyn. Res.,} 2001 \textbf{28}, 349.

     \bibitem[\protect\citeauthoryear{Hide and Titman}{1967}]{Hide67}
     Hide, R. and Titman, C.W.
     Detached shear layers in a rotating fluid,
     \textit{J.~Fluid Mech.,} 1967 \textbf{29}, 39-60.
     
     \bibitem[\protect\citeauthoryear{Ishihara and Kida}{2002}]{Ishihara02}
     Ishihara, N. and Kida, S. 
     Dynamo mechanism in a rotating spherical shell: competition between magnetic field and
     convection vortices,
     \textit{J. Fluid Mech.,} 2002 \textbf{465}, 1-32.
     
     \bibitem[\protect\citeauthoryear{Jones \textit{et al.}}{2000}] {Jones00}
     Jones, C.A., Soward, A.M. and Mussa, A.I. 
     The onset of thermal convection in a rapidly rotating sphere,
     \textit{J.~Fluid Mech.,} 2000 \textbf{405}, 157-179.
     
     \bibitem[\protect\citeauthoryear{Kageyama and Sato}{1997}]{Kageyama97}
     Kageyama, A. and Sato, T. 
     Generation mechanism of a dipole field by a magneto-hydrodynamic dynamo,
     \textit{Phys. Rev. E,} 1997 \textbf{55}, 4617-4626.
     
     \bibitem[\protect\citeauthoryear{Kim \textit{et al.}}{1999}]{Kim99}
     Kim, E., Hughes, D.W., Soward, A.M. 
     An investigation into high conductivity dynamo action driven by rotating convection,
     \textit{Geophys. Astrophys. Fluid Dyn.,} 1999 \textbf{91}, 303-332.   

     \bibitem[\protect\citeauthoryear{Morin and Dormy}{2004}]{Morin04}
     Morin, V. and Dormy, E. 
     Time dependent $\beta$-convection in rapidly rotating spherical shells,
     \textit{Phys. Fluids,} 2004 \textbf{16}, 1603. 
     
     \bibitem[\protect\citeauthoryear{M\"uller \textit{et al.}}{2004}] {Muller04}
     M\"uller, U., Stieglitz, R. and Horanyi, S.
     A two-scale hydromagnetic dynamo experiment,
     \textit{J.~Fluid Mech.,} 2004 \textbf{498}, 31-71.
             
     \bibitem[\protect\citeauthoryear{M\"uller \textit{et al.}}{2006}] {Muller06}
     M\"uller, U., Stieglitz, R. and Horanyi, S. 
     Experiments at a two-scale dynamo test facility,
     \textit{J.~Fluid Mech.,} 2006 \textbf{552}, 419-440.

     \bibitem[\protect\citeauthoryear{Nataf \textit{et al.}}{2006}]{Nataf06}
     Nataf H.-C., Alboussi{\`e}re T., Brito D., Cardin P., Gagniere N., Jault D., Masson J.-P. and
     Schmitt D. 
     Experimental study of super-rotation in a magnetostrophic spherical Couette flow,
     \textit{Geophys. Astrophys. Fluid Dyn.,} 2006 \textbf{100}, 281-298.   
     
     \bibitem[\protect\citeauthoryear{Olson \textit{et al.}}{1999}]{Olson99}
     Olson, P., Christensen, U. and Glatzmaier, G.A. 
     Numerical modeling of the geodynamo: Mechanisms of field generation and equilibration,
     \textit{J. Geophys. Res.,} 1999 \textbf{104}, 10383-10404.
     
     \bibitem[\protect\citeauthoryear{R\"adler \textit{et al.}}{2002a}]{Radler02a}
     R\"adler, K.-H., Rheinhardt, M., Apstein, E. and Fuchs, H. 
     On the mean-field theory of the Karlsruhe dynamo experiment I. Kinematic theory,
     \textit{Magnetohydrodynamics,} 2002 \textbf{38}, 41-71.
     

     \bibitem[\protect\citeauthoryear{Roberts}{1968}]{Roberts68}
     Roberts, P.H. 
     On the thermal instability of a rotating-fluid sphere containing heat sources,
     \textit{Philos. Trans. R. Soc. London Ser. A,} 1968 \textbf{263}, 93-117.    
     
	   \bibitem[\protect\citeauthoryear{Schaeffer and Cardin}{2005a}]{Schaeffer05a}
     Schaeffer, N. and Cardin, P. 
     Quasi-geostrophic model of the instabilities of the Stewartson layer,
     \textit{Phys. Fluids,} 2005 \textbf{17}, 104111.
     
 	   \bibitem[\protect\citeauthoryear{Schaeffer and Cardin}{2005b}]{Schaeffer05b}
     Schaeffer, N. and Cardin, P. 
     Rossby-wave turbulence in a rapidly rotating sphere,
     \textit{Nonlinear Processes in Geophysics,} 2005 \textbf{12}, 947-953.
     
 	   \bibitem[\protect\citeauthoryear{Schaeffer and Cardin}{2006}]{Schaeffer06}
     Schaeffer, N. and Cardin, P. 
     Quasi-geostrophic kinematic dynamos at low magnetic Prandtl number,
     \textit{Earth Planet. Sci. Lett.,} 2006 \textbf{245}, 595-604.
     
     \bibitem[\protect\citeauthoryear{Schmitt \textit{et al.}}{2008}]{Schmitt08}
     Schmitt D., Alboussiere T., Brito D., Cardin P., Gagnière N., Jault D. and Nataf H. 
     Rotating spherical Couette flow in a dipolar magnetic field: experimental study of
     magneto-inertial waves,
     \textit{J. Fluid Mech.,} 2008 \textbf{604}, 175-197.
          
     \bibitem[\protect\citeauthoryear{Schrinner \textit{et al.}}{2005}]{Schrinner05}
     Schrinner, M., R{\"a}dler, K.-H., Schmitt D., Rheinhardt M. and Christensen U.
     Mean-field view on rotating magnetoconvection and a geodynamo model,
     \textit{Astron. Nachr.,} 2005 \textbf{326}, 245-249.
     
     \bibitem[\protect\citeauthoryear{Schrinner \textit{et al.}}{2006}]{Schrinner06}
     Schrinner, M., R{\"a}dler, K.-H., Schmitt D., Rheinhardt M. and Christensen U. 
     Mean-field concept and direct numerical simulations of rotating magnetoconvection and the
     geodynamo,
     \textit{Geophys. Astrophys. Fluid Dyn.,} 2007 \textbf{101}, 81-116.
     
     \bibitem[\protect\citeauthoryear{Sommeria}{1986}]{Sommeria86}
     Sommeria, J. 
     Experimental study of the two-dimensional inverse energy cascade in a square box,
     \textit{J.~Fluid Mech.,} 1986 \textbf{170}, 139-168.
   
     \bibitem[\protect\citeauthoryear{Soward}{1977}]{Soward77}
     Soward, A.M. 
     On the finite amplitude thermal instability of a rapidly rotating fluid sphere,
     \textit{Geophys. Astrophys. Fluid Dyn.,} 1977 \textbf{9}, 19-74.
     
     \bibitem[\protect\citeauthoryear{Stefani and Gerbeth}{2005}]{Stefani05}
     Stefani, F. and Gerbeth, G. 
     Asymmetry polarity reversals, bimodal field distribution, and coherence resonance in a
     spherically          
     symmetric mean-field dynamo model,
     \textit{Phys. Rev. Lett.,} 2005 \textbf{94}, 184506. 
     
     \bibitem[\protect\citeauthoryear{Sumita and Olson}{2000}]{Sumita00}
     Sumita, I. and Olson, P. 
     Laboratory experiments on high Rayleigh number thermal convection in a rapidly rotating
     hemispherical 
     shell,
     \textit{Phys. Earth Planet. Inter.,} 2000 \textbf{117}, 153-170.
   
     \bibitem[\protect\citeauthoryear{Yano}{1992}]{Yano92}
     Yano, J.-I. 
     Asymptotic theory of thermal convection in rapidly rotating systems,
     \textit{J.~Fluid Mech.,} 1992 \textbf{243}, 103-131.

     \bibitem[\protect\citeauthoryear{Zhang}{1991}]{Zhang91}
     Zhang, K. 
     Convection in a rapidly rotating spherical shell at infinite Prandtl number; steadily
     drifting rolls,
     \textit{Phys. Earth Planet. Inter.,} 1991 \textbf{68}, 156-169.

     \bibitem[\protect\citeauthoryear{Zhang}{1992}]{Zhang92}
     Zhang, K.
     Spiralling columnar convection in rapidly rotating spherical fluid shells,
     \textit{J.~Fluid Mech.,} 1992 \textbf{236}, 535-556.
     
     \bibitem[\protect\citeauthoryear{Zhang and Jones}{1993}]{Zhang93}
     Zhang, K. and Jones, C.A. 
     The influence of Ekman boundary layers on rotating convection in spherical fluid shells,
     \textit{Geophys. Astrophys. Fluid Dyn.,} 1993 \textbf{71}, 145-162.
     
     \bibitem[\protect\citeauthoryear{Zhang}{1995}]{Zhang95}
     Zhang, K. 
     On coupling between the Poincar\'e euqation and the heat equation: non-slip boundary
     condition,
     \textit{J.~Fluid Mech.,} 1995 \textbf{284}, 239-256.
     
     \bibitem[\protect\citeauthoryear{Zhang \textit{et al.}}{2007}]{Zhang07}
     Zhang, K., Liao, X. and Busse, F.H. 
     Asymptotic solutions of convection in rapidly rotating non-slip spheres,
     \textit{J.~Fluid Mech.,} 2007 \textbf{578}, 371-380.
     
\end{thebibliography}
\end{document}